\documentclass[prl,twocolumn,superscriptaddress,floatfix,preprintnumbers]{revtex4-1}
\usepackage{graphics,amssymb,amsmath,epsfig}
\usepackage{graphicx}
\usepackage[usenames]{color}
\definecolor{goodgreen}{rgb}{0.1,0.5,0}
\definecolor{goodred}{rgb}{0.7,0,0}
\usepackage{pifont}
\usepackage{float}
\usepackage{multirow}
\usepackage{booktabs}
\usepackage[colorlinks,urlcolor=goodgreen,citecolor=blue,linkcolor=goodred]{hyperref}
\usepackage{bm}
\usepackage{mathtools}
\allowdisplaybreaks
\usepackage{calligra}
\usepackage{placeins}

\DeclareMathAlphabet{\mathcalligra}{T1}{calligra}{m}{n}
\DeclareFontShape{T1}{calligra}{m}{n}{<->s*[2.2]callig15}{}
\newcommand{\scriptr}{\mathcalligra{r}\,}

\renewcommand{\thefigure}{{\bf \arabic{figure}}}

\newcommand*{\citen}{}
\DeclareRobustCommand*{\citen}[1]{%
  \begingroup
    \romannumeral-`\x 
    \setcitestyle{numbers}%
    \cite{#1}%
  \endgroup
}

\begin{document}
\title{Defect-induced electronic smectic state at the surface of nematic materials}
\author{Aritra Lahiri}
\altaffiliation{Present address: Theoretical Physics 4, University of W\"urzburg, 97074 W\"urzburg, Germany}
\affiliation{School of Physics and Astronomy, University of Minnesota, Minneapolis, Minnesota 55455, USA}
\author{Avraham Klein}
\affiliation{Physics Department, Ariel University, Ariel 40700}
\author{Rafael M. Fernandes}
\email[]{rfernand@umn.edu}
\affiliation{School of Physics and Astronomy, University of Minnesota, Minneapolis, Minnesota 55455, USA}

\date{\today}

\begin{abstract}
  Due to the intertwining between electronic nematic and elastic degrees of freedom, lattice defects and structural inhomogeneities commonly found in crystals can have a significant impact on the electronic properties of nematic materials. Here, we show that defects commonly present at
the surface of crystals generally shift the wave-vector of the nematic instability to a non-zero value, resulting in an incommensurate electronic smectic phase. Such a smectic state onsets above the bulk nematic transition temperature and is localized near the surface of the sample. We argue that
this effect may explain not only recent observations of a modulated nematic phase in iron-based superconductors, but also several previous puzzling experiments that reported signatures consistent with nematic order before the onset of a bulk structural distortion.
\end{abstract}

\maketitle
Electronic nematicity has been observed in a wide range of systems,
including high-$T_{c}$ superconductors \citep{Fradkin2010,Lawler2010,Chu710},
heavy-fermion materials \citep{Okazaki2011,Ronning2017,Seo2020},
topological superconductors \citep{Hecker2018,Cho2020}, cold atoms
\citep{cold_atoms_nem}, and twisted moir\'e devices \citep{Cao2021,Rubio2020}.
Among those, iron-based superconductors (FeSC) have provided unique
insight into this quantum electronic state due to the nearly-universal
and unambiguous presence of nematic order and nematic fluctuations
in their phase diagrams \citep{Chu824,Chuang181,Chu710,Fernandes_2012,BOHMER201690,GALLAIS2016}.
Despite significant progress, essential questions remain unresolved,
related not only to the microscopic mechanisms of nematicity, but
also to its general phenomenology \citep{Fernandes2014}. For instance,
since early studies of FeSC, various probes in nominally unstrained
samples have reported signatures consistent with nematicity above
the nematic transition temperature $T_{\mathrm{nem}}$ established
by thermodynamic probes \citep{Kasahara2012,PhysRevLett.121.027001,doi:10.7566/JPSJ.84.043705,Rosenthal2014,PhysRevB.92.134521,PhysRevB.94.094509,PhysRevB.86.024519,PhysRevB.89.045101,Sonobe2018,PhysRevB.91.214503,PhysRevB.97.094515,PhysRevB.96.180502}.
More recently, experiments have found evidence for a spatially-modulated
nematic phase -- i.e. an electronic smectic phase \citep{Yuan2021,Li2017,Yim2018,Shimojima2021}.

The probes used in many of these experiments are particularly
sensitive to the surface, e.g. angle-resolved photo-emission spectroscopy
(ARPES) \citep{PhysRevB.89.045101,Sonobe2018,PhysRevB.91.214503},
scanning tunneling microscopy (STM) \citep{Rosenthal2014,Li2017,Yim2018,Yuan2021}, spatially resolved
photomodulation  \citep{PhysRevLett.121.027001},
and photo-emission electron microscopy (PEEM) \citep{Shimojima2021}. Moreover, the onset of these interesting phenomena does not usually
show typical phase-transition signatures in thermodynamic quantities,
such as specific heat \citep{PhysRevB.91.094512} and elasto-resistance
\citep{Chu710}. This suggests that both effects -- nematic manifestations
above $T_{\mathrm{nem}}$ and modulated nematic order -- may signal a surface nematic transition at higher temperatures than the bulk one \citep{PhysRevLett.121.027001}, reminiscent of the so-called extraordinary transition \citep{Cardy1996}. The key question is whether
a surface nematic transition is particular to some FeSC compounds
or a more general phenomenological property of nematic compounds.

\begin{figure}
\centering \includegraphics[clip,width=0.7\columnwidth]{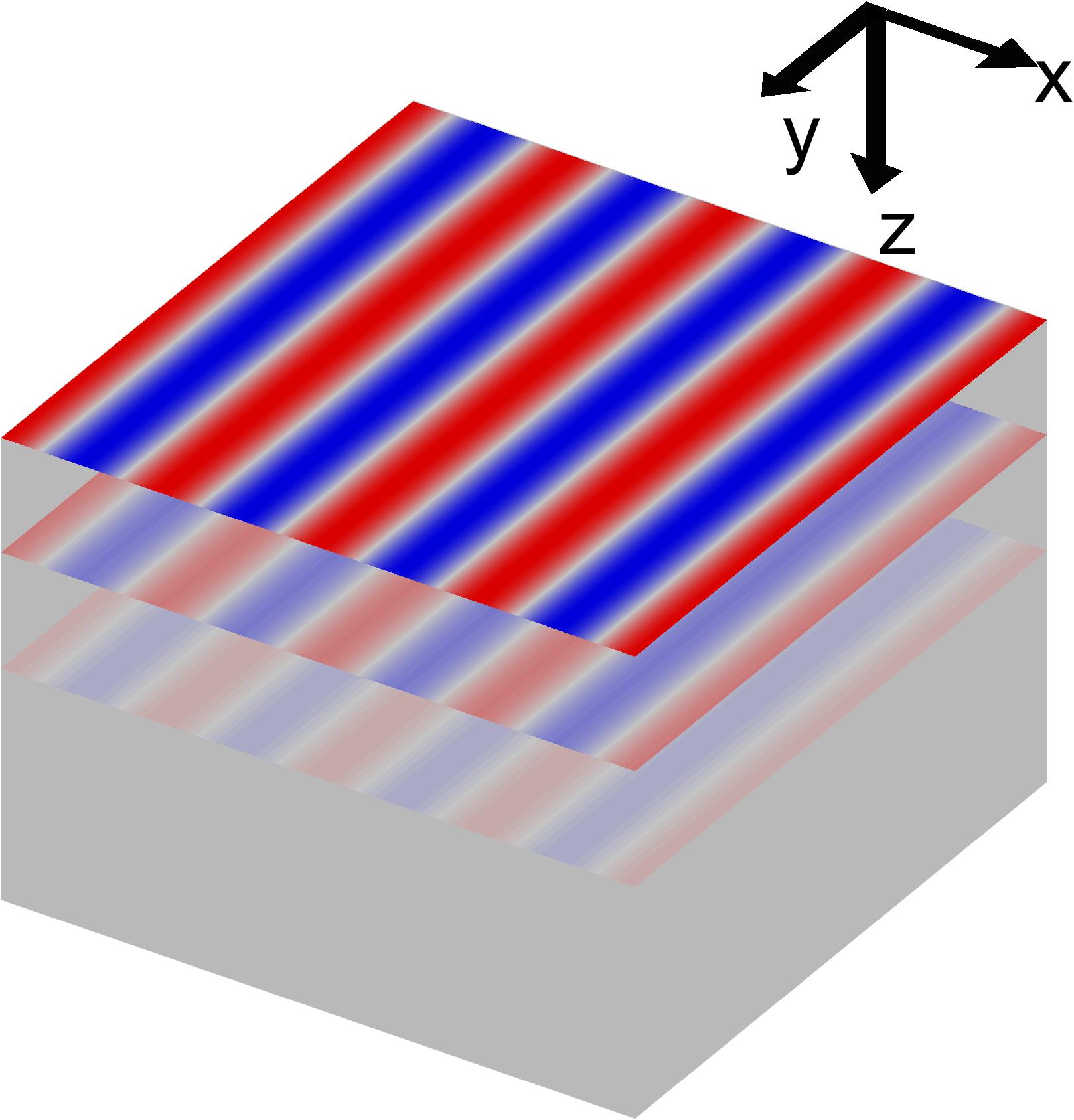}\llap{\raisebox{0.05\hsize}{\includegraphics[width=0.45\columnwidth]{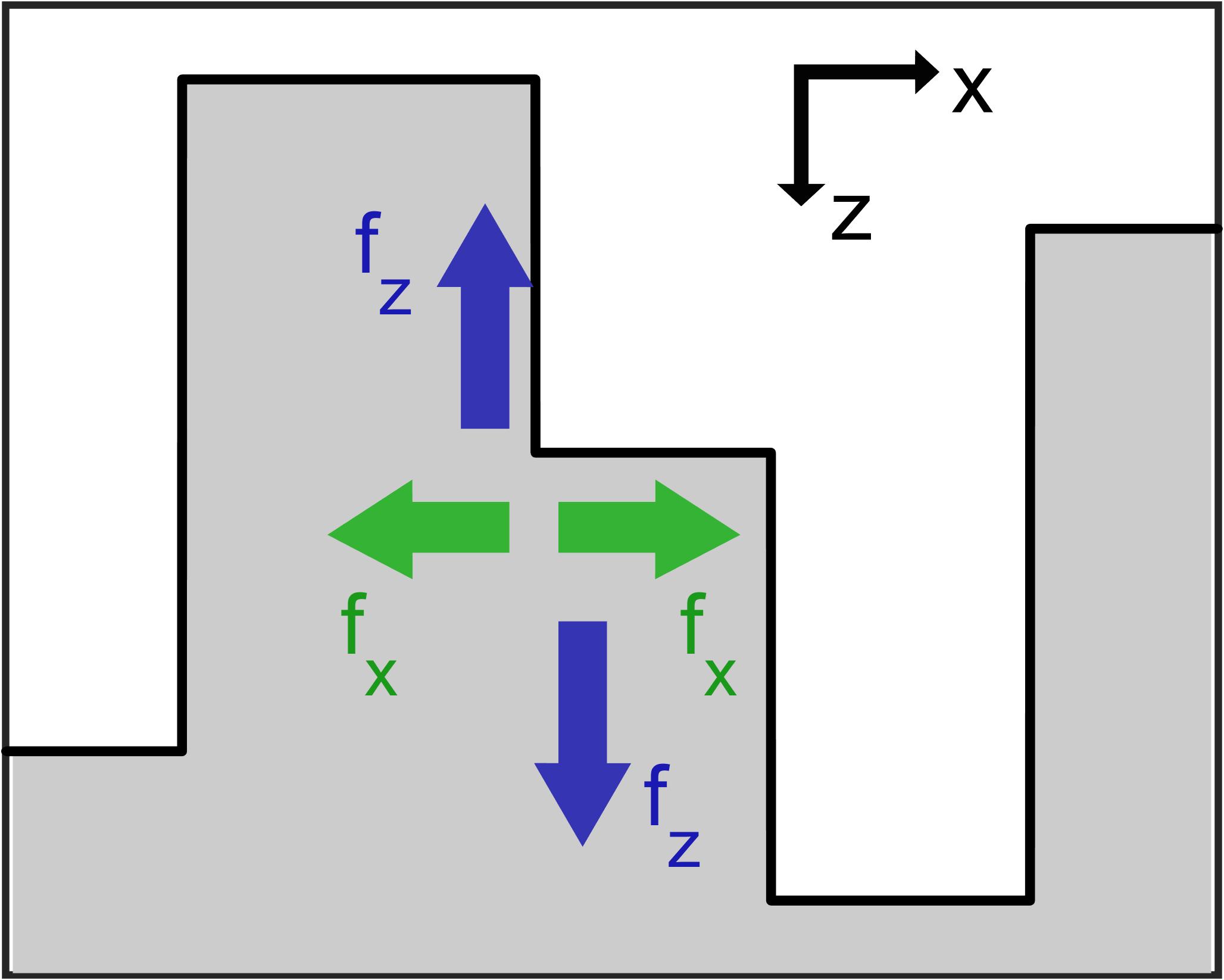}}}
\caption{(color online) Schematic illustration of the surface smectic state,
shown here as a modulated nematic order parameter that quickly decays
in the bulk of the sample (gray). Red (blue) regions denote a $B_{1g}$
nematic order parameter that selects the $x$ ($y$) axis of a tetragonal
crystal. The inset illustrates the dipolar forces induced by a surface
step. It also presents a cross-section of the
sample (gray) with aligned steps of random heights/strengths
oriented parallel to the $y-$axis.}
\label{fig_smec}
\end{figure}

While a purely electronic mechanism was previously invoked to explain
surface nematicity \citep{Koshelev2016}, in this Letter we focus
on the role of the elastic degrees of freedom. The nemato-elastic
coupling $g$ is known to significantly impact
the nematic state, particularly in FeSC \citep{PhysRevLett.105.157003,doi:10.1143/JPSJ.80.073702,doi:10.1143/JPSJ.81.024604,PhysRevLett.111.137001,BOHMER201690,PhysRevLett.124.157001,Chibani2021,GALLAIS2016}.
For instance, coupling to elastic fluctuations (acoustic
phonons) renders the
nematic transition mean-field like \citep{Qi2009,Schmalian2016,Paul_Garst,Carvalho2019,PhysRevLett.124.157001}, whereas
intrinsic random strain fosters behaviors associated with the
random-field Ising-model \citep{Carlson2006,Kuo2016,Wiecki2021}. Here, we show
that defects commonly found in the surfaces of crystals, such as steps separating terrace domains, promote an electronic smectic
state localized near the surface and that onsets at a temperature $T_{\mathrm{smc}} >T_{\mathrm{nem}}$
(see Fig. \ref{fig_smec}). The smectic state survives down to a temperature $T_{\mathrm{smc-nem}}$, which decreases as the sample thickness is reduced, at which point a homogeneous nematic phase takes over. Our results establish a hitherto unexplored facet of electronic nematic phases in elastic media, which we argue can explain the intriguing observation of Ref. \citep{Shimojima2021} of a mesoscopic nematic wave in FeSC.

To understand why defects induce a
surface transition, note that elastic fluctuations increase the
nematic transition temperature $T_{\mathrm{nem}}$ from its bare purely-electronic value $T_{\mathrm{nem}}^{(0)}$. In a clean system, some
of the elastic modes are expected to be frozen near the surface, resulting in  $T_{\mathrm{nem}}^{(\mathrm{surface})}<T_{\mathrm{nem}}$
\citep{athesis}. However, the fact that the exposed
surface is more disordered than the bulk changes this picture dramatically.
To see this, consider a random distribution of defects, such as vacancies
and dislocations, on the surface of a crystal whose bulk is clean.
Defects locally induce large strains that decay
slowly with distance \citep{LIFSHITZ198687}. Since they are concentrated at the surface, they rapidly screen each other as one moves deeper into the bulk. However, near the surface, they do not screen efficiently, causing not only an enhancement of $T_{\mathrm{nem}}$ at the surface, but also creating a ``speckle'' pattern in the nematic fluctuation spectrum, with typical spot size set by the algebraic strain correlations rather than by the defect density. This disorder-induced pattern imposes a preferred wavelength for the condensation of the nematic order parameter, driving the formation of an electronic smectic state.

To derive these results, we solve a Ginzburg-Landau model of a generic nematic order parameter coupled to elastic strain induced by simple types of surface quenched disorder, such as steps and anisotropic point defects. We find that the defect distribution induces a non-local effective potential for the nematic order parameter. After averaging over defect realizations, the minimum of the resulting nematic free energy appears at a higher temperature $T_{\mathrm{smc}}=T_{\mathrm{nem}}+\Delta T_{\mathrm{smc}}$
(with $\Delta T_{\mathrm{smc}}>0$) and at a non-zero wave-vector
$q_{\mathrm{smc}}$, resulting in an electronic smectic phase. In
terms of the disorder strength $\sigma^{2}$, we find
\begin{equation}
q_{\mathrm{smc}}\propto g^{2}\sigma^{2},\quad\Delta T_{\mathrm{smc}}\propto q_{\mathrm{smc}}^{2},\label{eq:T-s}
\end{equation}
The smectic order parameter is inhomogeneous and localized
at the the surface, decaying exponentially into the bulk with
a penetration depth $\propto1/q_{\text{smc}}$. Eventually, below $T_{\mathrm{smc-nem}}$, which is lower than the bulk nematic transition temperature $T_{\text{nem}}$, the smectic solution becomes unfavorable and the uniform $q=0$ nematic state is established throughout the sample.

\textit{Surface step disorder and induced strain.--} To elucidate our results, we consider an Ising-nematic order parameter $\eta$ that breaks the equivalence between the $x$ and $y$ directions of a crystal (i.e. it transforms as the $B_{1g}$ irreducible representation of the tetragonal group). In the presence of strain, the nematic action is given by:
\begin{equation}
S=\int_{\bm{r}} \left[\bigg(r_0\frac{T-T_{\mathrm{nem}}}{2T_{\mathrm{nem}}^{(0)}}\bigg)\eta_{\bm{r}}^{2}+\frac{b_{\mu}}{2}(\partial_{\mu}\eta_{\bm{r}})^{2}-g\varepsilon_{\bm{r}}^{B_{1g}}\eta_{\bm{r}} + \frac{u_\eta}{4} \eta_{\bm{r}}^4 \right]\label{S_nem}
\end{equation}
where repeated indices are implicitly summed; $b_{x}=b_{y}=b_{\parallel}$
and $b_{z}$ are the nematic stiffness coefficients; $u_\eta > 0$ is the quartic coefficient; $r_0$ is of the order of the Fermi energy (action has dimensions of energy); $\varepsilon^{B_{1g}}\equiv(\varepsilon_{xx}-\varepsilon_{yy})/\sqrt{2}$ is the $B_{1g}$ shear strain, which acts as a conjugate field to the nematic order parameter; and $T_{\mathrm{nem}}^{(0)},T_{\mathrm{nem}}$ are the nematic transition temperatures without and with the enhancement from elastic fluctuations. For a clean unstrained crystal, $\varepsilon_{\bm{r}}^{B_{1g}}$ is only present as a fluctuating field whose properties are determined by the crystal's elastic constants. However, for a crystal with quenched disorder, a static slow-decaying strain $\varepsilon_{\bm{r}}^{B_{1g}}$ is generated by the various types of defects. In both cases, an effective nematic potential emerges in the action due to either thermal fluctuations or average over disorder configurations. While the former scenario has been widely studied \citep{Qi2009,Schmalian2016,Paul_Garst,Carvalho2019}, the latter has received much less attention \citep{Nie2014,Cui2018}.

A crystal with an exposed surface can be modeled by an isotropic
elastic half-space ($z\geq0$) with Young's modulus
$E$ and Poisson ratio $\nu$. Each type of surface defect generates
a characteristic dipolar local force, which in turn can be used to calculate $\varepsilon_{\bm{r}}^{B_{1g}}$ via standard methods \citep{MarchenkoParshin,PhysRevB.53.11120,PhysRevB.49.13848,doi:https://doi.org/10.1002/352760667X.ch9,BACON198051,CLOUET201849,Teodosiu1982}.
Here, we consider idealized infinite step defects parallel to the
$y-$axis, as shown in Fig. \ref{fig_smec} (we consider point defects in the Supplementary material (SM)).
A single step at $x=x'$ is parametrized by the force density $f_{\mu}=h_{\mu}[\partial_{x}\delta(x-x')]\delta(z)$, where $\delta(z)$ is the Dirac delta function, the force $h_{\mu}$ characterizes the strength of the defect, and $\mu=x,z$. For simplicity, we consider steps that create forces along the $z-$axis only, i.e. $h_{x}=0$ and $h_{z}\neq0$. The lattice displacement created by a single step is given by $u_{\mu}=h_{\nu}\partial_{x}G_{\mu\nu}(x-x',z)$, where $G_{\mu\nu}$ is the Green's function for an infinite line-force along the $y-$axis in half-space \citep{LIFSHITZ198687}. The $B_{1g}$ strain $\varepsilon_{\bm{r}-\bm{r}'}^{B_{1g}}$ generated by a single defect is \citep{LIFSHITZ198687}
\begin{equation}
\varepsilon_{\bm{r}-\bm{r}'}^{B_{1g}} =\frac{-4(1+\nu)h_{z}}{\sqrt{2}\pi E}\bigg[\frac{(\nu-1)\,\delta x^{3}z+(\nu+1)\,\delta x\,z^{3}}{\big(\delta x^{2}+z^{2}\big)^{3}}\bigg],
\end{equation}
where $\delta x= x-x'$. A distribution
of such steps at random positions $x=x_{j}$ and with
random strength $h_{z,j}$ results in the net $B_{1g}$ strain $\varepsilon_{\bm{r}}^{B_{1g}}=\sum_{j}h_{j}\partial_x^2G_{xz}(x-x',z-0)\equiv\sum_{j}h_{j}\bar{\varepsilon}_{\bm{r}-\bm{r}_{j}}^{B_{1g}}$. The nematic action (\ref{S_nem}) for the finite crystal with dimensions $L_{x}=L_{y}=L_{\parallel}$ and $L_{z}=L\ll L_{\parallel}$
becomes: 
\begin{equation}
\begin{split}S=& L_\parallel\int_{-\frac{L_{\parallel}}{2}}^{\frac{L_{\parallel}}{2}}dx\int_{0}^{L}dz\bigg[\bigg(r_0\frac{T-T_{\mathrm{nem}}}{2T_{\mathrm{nem}}^{(0)}}\bigg)\eta_{x,z}^{2}+\frac{b_{\parallel}}{2}(\partial_{x}\eta_{x,z})^{2}\\
 & +\frac{b}{2}(\partial_{z}\eta_{x,z})^{2}+\frac{u_\eta}{4} \eta_{x,z}^4-g\int_{-\frac{L_{\parallel}}{2}}^{\frac{L_{\parallel}}{2}}dx'\rho_{x'}\bar{\varepsilon}_{x-x',z}^{B_{1g}}\eta_{x,z}\bigg],\end{split}\label{S_deffect}
\end{equation}
where we defined $\rho_{x}=\sum_{j}h_{j}\delta(x-x_{j})$.

\textit{Effective nematic potential and smectic state.---} For a random distribution of steps, $\langle h_{j}h_{j'}\rangle=\sigma^{2}\delta_{j,j'}$, the step density $\rho_{x}$ follows a Gaussian distribution with variance $\sigma^{2}(N_{\text{step}}/L_{\parallel})(a_{\parallel}/L_\xi)$, where $N_{\text{step}}$ is the number of steps, $a_\parallel$ is the in-plane lattice constant, and $L_\xi$ is a length scale larger than $a_\parallel$ but smaller than the nematic correlation length. Integrating out the step density in Eq. \eqref{S_deffect} (equivalent to the standard procedure of averaging over quenched disorder \citep{dedominicis_giardina_2006,dotsenko_2000,PhysRevLett.35.1399,PhysRevLett.37.1364,PhysRevLett.37.944,M_zard_1992,Young_1977}) generates a new quadratic term in the nematic action: 
\begin{equation}
S_{d}=L_\parallel^2\int_{0}^{L}dz\,dz'\sum_{q_{x}}V_{q_{x},z,z'}\eta_{q_{x},z}^{*}\eta_{q_{x},z'}\label{eq:S-d}
\end{equation}
with an effective potential experienced by the nematic order parameter
\begin{equation}
\begin{split}V_{q_{x},z,z'}= & -\frac{(g\sigma)^{2}\beta}{2}\,\mathrm{e}^{-|q_{x}|(|z|+|z'|)}\\
 & \times q_{x}^{2}[|q_{x}||z|+2\nu-1][|q_{x}||z'|+2\nu-1].
\end{split}
\label{reta}
\end{equation}
Here, $\beta=[(1+\nu)/(\sqrt{2}E)]^{2}N_{\text{step}}(L_\xi/a_\parallel)$ and $\eta_{q_{x},z}=(1/L_{\parallel})\int_{x}\eta_{x,z}e^{-iq_{x}x}$. The potential $V_{q_{x},z,z'}$ is non-local, depending on both $z$ and $z'$. Moreover, it vanishes quadratically as $q_{x} \rightarrow 0$ and exponentially as $z,\,z'\rightarrow\infty$ or $q_{x} \rightarrow \infty$. Thus, the potential has a negative-valued minimum at a non-zero $q_x$ and is significant only near the surface. These features are a consequence of the algebraic decay of the strain fields generated by defects, rather than the type of defects (see SM). 

\begin{figure*}
\includegraphics[width=7in]{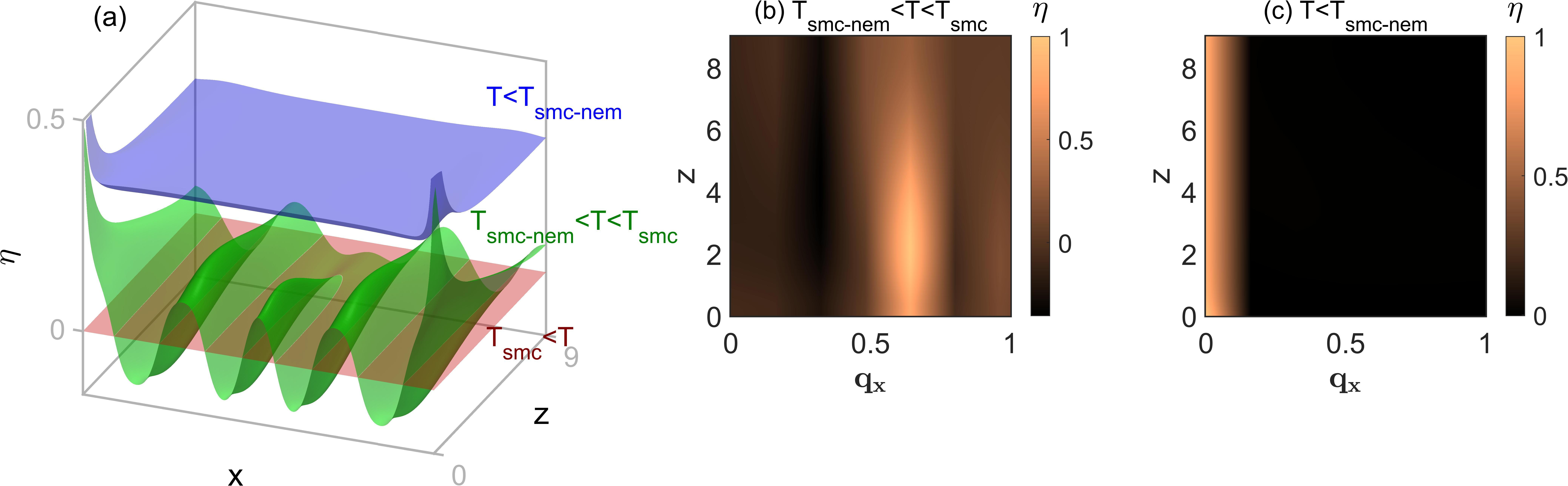}
\caption{(color online) (a) Spatial profile of the nematic order parameter
  $\eta_{x,z}$
for three representative temperatures,
obtained from the numerical solution of the saddle-point equation
(\ref{sadrealstep}). For $T>T_{\mathrm{smc}}$, the nematic order parameter is effectively zero everywhere. As temperature is lowered towards $T_{\mathrm{smc-nem}}<T<T_{\mathrm{smc}}$, $\eta_{x,z}$ displays a sinusoidal $x$-dependence characterized by a single smectic wave-vector $q_{\mathrm{smc}}$ [panel (b)]. Below the bulk nematic transition $T<T_{\mathrm{smc-nem}}$, a uniform nematic state emerges with zero wave-vector [panel (c)]. The enhancement of $\eta_{x,z}$ at the corners is an artifact of the boundary conditions. The profile of the nematic order parameter $\eta_{q_{x},z}$ in Fourier space is shown in panels (b) (for $T_{\mathrm{smc-nem}}<T<T_{\mathrm{smc}}$) and (c) (for $T<T_{\mathrm{smc-nem}}$). The parameters used are (in arbitrary units): $r_0=1$, $b=0.5$, $b_\parallel=0.25$, $\nu=0.495$, $L_\parallel=44$, $L=9$, $(g\sigma)^2\beta/2=1$, and $u_\eta=5$. In panels (b) and (c), the nematic fields were normalized.}
\label{fig_profile} 
\end{figure*}

While the defect-generated potential in Eq. \eqref{reta} is minimized by
$q_{x} \neq 0$, the nematic stiffness term $b_{\parallel}q_{x}^{2}$ in
Eq. \eqref{S_deffect} favors a uniform $q_{x}=0$ state.
This competition causes the nematic instability to take place at a nonzero
wave-vector $q_{x}$, resulting in an electronic smectic state. This effect is restricted to the vicinity of the surface due to the exponential suppression of $V_{q_{x},z,z'}$ with $|z|$. This can be more clearly seen by an
approximate analytical solution of the problem. Re-expressing $V_{q_{x},z,z'}$ in terms of $\bar{z}=(z+z')/2$ and $\delta z=z-z'$,
$V_{q_{x},\bar{z},\delta z}$ is peaked at $\bar{z}\sim1/|q_{x}|$
and $\delta z=0$. Assuming that $\eta_{q_{x},z}$ varies slowly
near the surface over a depth $L_s \sim 1/|q_{x}|$, before eventually decaying exponentially away from the surface, the action \eqref{eq:S-d} becomes:
\begin{equation}
\begin{split}S_{d}= & L_\parallel^2\int_{\bar{z}=0}^{L}\int_{\delta z=-\frac{L_{s}}{2}}^{\frac{L_{s}}{2}}\sum_{q_{x}}V_{q_{x},\bar{z},\delta z}|\eta_{q_{x},0}|^{2},\\
\approx & -L_\parallel^2 L_s\sum_{q_{x}}\frac{(g\sigma)^{2}\beta\big[(\nu-\frac{1}{2})^{2}+\nu^{2}\big]}{2}\,|q_{x}||\eta_{q_{x},0}|^{2}.
\end{split}
\label{eq:S-d-q_x-int}
\end{equation}

In the regime of vanishing $z$-component stiffness $b\to0$, the quadratic
part of the action (\ref{S_deffect}), $S^{(2)}$, is given by:

\begin{equation}
S^{(2)}\approx L_\parallel^2 L_s\sum_{q_{x}}\left[\bigg(r_0\frac{T-T_{\mathrm{nem}}}{2T_{\mathrm{nem}}^{(0)}}\bigg)+\frac{b_{\parallel}q_{x}^{2}}{2}\right]|\eta_{q_{x},0}|^{2}\label{S_2}
\end{equation}

Minimizing the full action $S_{d}+S^{(2)}$ with respect to $q_{x}$
gives a finite smectic wave-vector $q_{\mathrm{smc}}=(g\sigma)^{2}\beta\big[(\nu-\frac{1}{2})^{2}+\nu^{2}\big]/2b_{\parallel}$
and an enhanced smectic transition temperature $T_{\mathrm{smc}}=T_{\mathrm{nem}}+(T_{\mathrm{nem}}^{(0)}/r_0)b_{\parallel}q_{\mathrm{smc}}^{2}$, consistent with Eq. \eqref{eq:T-s}. The actual spatial profile of $\eta_{x,z}$ and the precise $q_{\mathrm{smc}}$ and $T_{\mathrm{smc}}$ can be obtained by solving the saddle-point equation in real space, 
\begin{align}
\bigg[r_0\frac{T-T_{\mathrm{nem}}}{T_{\mathrm{nem}}^{(0)}}-b\partial_{z}^{2}-b_{\parallel}\partial_{x}^{2}\bigg]\eta_{x,z}+u_{\eta}\eta_{x,z}^{3}\nonumber\\
+\frac{1}{L_\parallel}\int_{0}^{L}dz'\int_{-\frac{L_{\parallel}}{2}}^{\frac{L_{\parallel}}{2}}dx'\,V_{x-x',z,z'}\eta_{x',z'} & =0\label{sadrealstep} 
\end{align}
where $V_{\delta x,z,z'}$ is the inverse Fourier transform of $V_{q_{x},z,z'}$ (see SM whose asymptotic behavior
is:
\begin{equation}
\begin{split}V_{\delta x,z,z'}\sim\left\{ \begin{array}{ll}
-(z+z')^{-3} & ,\ |\delta x|\ll z,z'\\
+(z+z')(\delta x)^{-4} & ,\ |\delta x|\gg z,z'
\end{array}\right..\end{split}
\end{equation}

Therefore, as a function of $\delta x/(z+z')$, $V_{\delta x,z,z'}$
has a negative central trough at $\delta x=0$, crosses zero at $\delta x\sim z+z'$, and then remains positive as it decays algebraically. The sign change
in real-space means that the effective potential favors
an oscillatory $\eta_{x}$ solution.

The numerical solution of Eq. \eqref{sadrealstep}, shown in Fig.
\ref{fig_profile}(a), confirms the main
results of our analytical approximation. The quartic term of the nematic
action {\eqref{S_deffect}} stabilizes a single smectic wave-vector over the entire temperature range $T_{\mathrm{smc-nem}}<T<T_{\mathrm{smc}}$, as it acts as a repulsive biquadratic interaction $u_{\eta}|\eta_{q_{x}}|^{2}|\eta_{q_{x}'\neq q_{x}}|^{2}$ between states with different wave-vectors. Consequently, only the smectic wave-vector corresponding to the highest critical temperature develops. For the same reason, in a fully 3D crystal with $L \gg 1/q_{\text{smc}}$, the uniform bulk nematic phase is preferred for $T<T_{\mathrm{nem}}$, as its free-energy gain scales extensively with the system size. However, for smaller values of $L$ comparable to $1/q_{\text{smc}}$, the smectic free-energy can compete with the bulk nematic one. Consequently, the smectic-nematic transition is pushed to a lower temperature $T_{\text{smc-nem}}<T_{\text{nem}}$, which decreases with decreasing sample thickness. Figs. \ref{fig_profile}(b)-(c) show the corresponding profile of $\eta_{q_{x},z}$ in momentum space, highlighting the change in wave-vector above and below $T_{\mathrm{smc-nem}}$.

\begin{figure}
\includegraphics[width=3.4in]{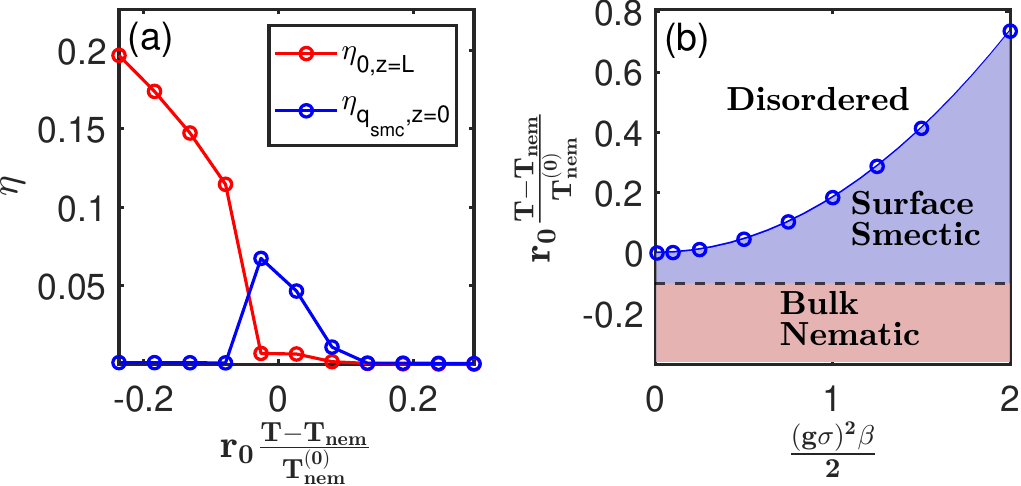}
\caption{(a) Temperature dependence of the uniform nematic (red, approximated by $\eta_{q_x=0, z=L}$) and surface smectic (blue,  approximated by $\eta_{q_x=q_{\mathrm{smc}}\approx0.57, z=0}$) order parameters, numerically obtained by solving Eq. \eqref{sadrealstep} (same parameters as Fig.\ref{fig_profile}). (b) Phase diagram as a function of the effective disorder strength $\frac{(g\sigma)^{2}\beta}{2}$ and the reduced temperature. The smectic critical temperature (blue circles), found to vary quadratically with $\frac{(g\sigma)^{2}\beta}{2}$, was obtained from the linearized saddle-point equation \eqref{sadrealstep} in momentum space (see SM). Due to the finite sample thickness, the bulk nematic phase onsets at $T=T_{\mathrm{smc-nem}}<T_{\text{nem}}$.}
\label{fig_phase_diagram} 
\end{figure}

The temperature dependence of the uniform nematic and smectic order
parameters is shown in Fig. \ref{fig_phase_diagram}(a). The continuous onset of surface smectic order is evident, eventually dropping discontinuously to zero, concomitant to the onset of uniform nematic order. Fig. \ref{fig_phase_diagram}(b) shows the numerically obtained phase diagram as a function of increasing defect disorder strength $\sigma^{2}$.

\textit{Discussion.---} The mechanism unveiled in this work for the emergence
of a surface electronic smectic state above the onset of bulk electronic
nematicity is rather general, as it relies solely on the existence
of defects commonly observed at crystal surfaces. While here we focused
on steps, other defects with nonzero dipolar elastic moments
are expected to promote a similar behavior, since they also generate
algebraically-decaying strain fields that are poorly screened at the
surface (see SM) \citep{Hameed2021,PhysRevB.104.024511}. Our result unearths yet another aspect of the rich phenomenology of electronic nematicity caused by the coupling
to the elastic degrees of freedom.

The impact of the effect we found on a given nematic system depends on the disorder strength $\sigma$ and on  the nemato-elastic coupling $g$, as shown in the phase diagram of Fig. \ref{fig_phase_diagram}(b). FeSC stand out as compounds with strongly coupled nematic and elastic degrees of freedom, as manifested by, e.g., the large orthorhombic distortion seen in the nematic phase \citep{Avci2012}. In contrast, in other tetragonal correlated systems that display nematic tendencies, such as Hg-based cuprates \citep{Murayama2019} and heavy-fermion systems \citep{Ronning2017,Seo2020}, a lattice distortion is difficult to be resolved experimentally. The potentially large
period $2\pi/q_{\mathrm{smc}}$ of the smectic state may explain why certain surface-sensitive probes, such as ARPES and STM, observe signatures
consistent with nematic order above the temperature where a bulk orthorhombic distortion onsets. Among the various experimental findings that have indicated the existence of a smectic phase in FeSC \citep{Yuan2021,Li2017,Yim2018,Shimojima2021}, the PEEM data reported in Ref. \citep{Shimojima2021} provides the most straightforward platform to perform comparisons with our theory and extract relevant physical estimates. That work found a sinusoidal modulation of the nematic order parameter with a long and material-dependent period. Moreover, when Fourier-transformed to momentum space (see the SM), the PEEM data, available in \citep{shimojima_takahiro_2021_4885407}, displays a distinctive speckle pattern corresponding to a spot of size $q_{\text{smc}}$, reminiscent of our theoretically calculated nematic potential $V_{q_{x},z,z'}$. As shown in detail in the SM, combining the experimental results of Ref. \citep{Shimojima2021} with our theoretical model, we find two interesting results: (i) The size of a typical region of parallel stripes, observed in that work, is small enough that a homogeneous nematic phase may not be stabilized at $T_{\mathrm{nem}}$.  (ii) The characteristic energy scale per defect is of the order $E_d \sim 100 \mu$eV. This scale is much smaller than both the Fermi energy and the bulk nematic orbital order energy splitting observed in FeSe. These results highlight that the smectic order is not the result of a particular defect distribution with fine-tuned disorder strength, but of the subtle effects of the long-range strain generated by the defect distribution.  {\nocite{shimojima_takahiro_2021_4885407,PhysRevResearch.2.013336,PhysRevB.96.121103,PhysRevLett.124.157001,CHANDRA20102072,PhysRevB.86.075123,PhysRevB.55.4737,MULLER2004157}}
 
\begin{acknowledgments}
We acknowledge fruitful discussions with B. Davidovitch, D. Pelc,
M. Greven, and J. Schmalian. This work was supported by the U. S.
Department of Energy, Office of Science, Basic Energy Sciences, Materials
Sciences and Engineering Division, under Award No. DE-SC0020045 (R.M.F.).
A.K. and R.M.F. acknowledge the hospitality of KITP at UCSB, where
part of the work was conducted. The research at KITP is supported
by the National Science Foundation under Grant No. NSF PHY-1748958. 
\end{acknowledgments}

\bibliographystyle{apsrev}
\bibliography{bibl}

\begin{thebibliography}{77}
\expandafter\ifx\csname natexlab\endcsname\relax\def\natexlab#1{#1}\fi
\expandafter\ifx\csname bibnamefont\endcsname\relax
  \def\bibnamefont#1{#1}\fi
\expandafter\ifx\csname bibfnamefont\endcsname\relax
  \def\bibfnamefont#1{#1}\fi
\expandafter\ifx\csname citenamefont\endcsname\relax
  \def\citenamefont#1{#1}\fi
\expandafter\ifx\csname url\endcsname\relax
  \def\url#1{\texttt{#1}}\fi
\expandafter\ifx\csname urlprefix\endcsname\relax\def\urlprefix{URL }\fi
\providecommand{\bibinfo}[2]{#2}
\providecommand{\eprint}[2][]{\url{#2}}

\bibitem[{\citenamefont{Fradkin et~al.}(2010)\citenamefont{Fradkin, Kivelson,
  Lawler, Eisenstein, and Mackenzie}}]{Fradkin2010}
\bibinfo{author}{\bibfnamefont{E.}~\bibnamefont{Fradkin}},
  \bibinfo{author}{\bibfnamefont{S.~A.} \bibnamefont{Kivelson}},
  \bibinfo{author}{\bibfnamefont{M.~J.} \bibnamefont{Lawler}},
  \bibinfo{author}{\bibfnamefont{J.~P.} \bibnamefont{Eisenstein}},
  \bibnamefont{and} \bibinfo{author}{\bibfnamefont{A.~P.}
  \bibnamefont{Mackenzie}}, \bibinfo{journal}{Annual Review of Condensed Matter
  Physics} \textbf{\bibinfo{volume}{1}}, \bibinfo{pages}{153}
  (\bibinfo{year}{2010}).

\bibitem[{\citenamefont{Lawler et~al.}(2010)\citenamefont{Lawler, Fujita, Lee,
  Schmidt, Kohsaka, Kim, Eisaki, Uchida, Davis, Sethna et~al.}}]{Lawler2010}
\bibinfo{author}{\bibfnamefont{M.~J.} \bibnamefont{Lawler}},
  \bibinfo{author}{\bibfnamefont{K.}~\bibnamefont{Fujita}},
  \bibinfo{author}{\bibfnamefont{J.}~\bibnamefont{Lee}},
  \bibinfo{author}{\bibfnamefont{A.~R.} \bibnamefont{Schmidt}},
  \bibinfo{author}{\bibfnamefont{Y.}~\bibnamefont{Kohsaka}},
  \bibinfo{author}{\bibfnamefont{C.~K.} \bibnamefont{Kim}},
  \bibinfo{author}{\bibfnamefont{H.}~\bibnamefont{Eisaki}},
  \bibinfo{author}{\bibfnamefont{S.}~\bibnamefont{Uchida}},
  \bibinfo{author}{\bibfnamefont{J.~C.} \bibnamefont{Davis}},
  \bibinfo{author}{\bibfnamefont{J.~P.} \bibnamefont{Sethna}},
  \bibnamefont{et~al.}, \bibinfo{journal}{Nature}
  \textbf{\bibinfo{volume}{466}}, \bibinfo{pages}{347} (\bibinfo{year}{2010}).

\bibitem[{\citenamefont{Chu et~al.}(2012)\citenamefont{Chu, Kuo, Analytis, and
  Fisher}}]{Chu710}
\bibinfo{author}{\bibfnamefont{J.-H.} \bibnamefont{Chu}},
  \bibinfo{author}{\bibfnamefont{H.-H.} \bibnamefont{Kuo}},
  \bibinfo{author}{\bibfnamefont{J.~G.} \bibnamefont{Analytis}},
  \bibnamefont{and} \bibinfo{author}{\bibfnamefont{I.~R.}
  \bibnamefont{Fisher}}, \bibinfo{journal}{Science}
  \textbf{\bibinfo{volume}{337}}, \bibinfo{pages}{710} (\bibinfo{year}{2012}).

\bibitem[{\citenamefont{Okazaki et~al.}(2011)\citenamefont{Okazaki, Shibauchi,
  Shi, Haga, Matsuda, Yamamoto, Onuki, Ikeda, and Matsuda}}]{Okazaki2011}
\bibinfo{author}{\bibfnamefont{R.}~\bibnamefont{Okazaki}},
  \bibinfo{author}{\bibfnamefont{T.}~\bibnamefont{Shibauchi}},
  \bibinfo{author}{\bibfnamefont{H.~J.} \bibnamefont{Shi}},
  \bibinfo{author}{\bibfnamefont{Y.}~\bibnamefont{Haga}},
  \bibinfo{author}{\bibfnamefont{T.~D.} \bibnamefont{Matsuda}},
  \bibinfo{author}{\bibfnamefont{E.}~\bibnamefont{Yamamoto}},
  \bibinfo{author}{\bibfnamefont{Y.}~\bibnamefont{Onuki}},
  \bibinfo{author}{\bibfnamefont{H.}~\bibnamefont{Ikeda}}, \bibnamefont{and}
  \bibinfo{author}{\bibfnamefont{Y.}~\bibnamefont{Matsuda}},
  \bibinfo{journal}{Science} \textbf{\bibinfo{volume}{331}},
  \bibinfo{pages}{439} (\bibinfo{year}{2011}).

\bibitem[{\citenamefont{Ronning et~al.}(2017)\citenamefont{Ronning, Helm,
  Shirer, Bachmann, Balicas, Chan, Ramshaw, McDonald, Balakirev, Jaime
  et~al.}}]{Ronning2017}
\bibinfo{author}{\bibfnamefont{F.}~\bibnamefont{Ronning}},
  \bibinfo{author}{\bibfnamefont{T.}~\bibnamefont{Helm}},
  \bibinfo{author}{\bibfnamefont{K.~R.} \bibnamefont{Shirer}},
  \bibinfo{author}{\bibfnamefont{M.~D.} \bibnamefont{Bachmann}},
  \bibinfo{author}{\bibfnamefont{L.}~\bibnamefont{Balicas}},
  \bibinfo{author}{\bibfnamefont{M.~K.} \bibnamefont{Chan}},
  \bibinfo{author}{\bibfnamefont{B.~J.} \bibnamefont{Ramshaw}},
  \bibinfo{author}{\bibfnamefont{R.~D.} \bibnamefont{McDonald}},
  \bibinfo{author}{\bibfnamefont{F.~F.} \bibnamefont{Balakirev}},
  \bibinfo{author}{\bibfnamefont{M.}~\bibnamefont{Jaime}},
  \bibnamefont{et~al.}, \bibinfo{journal}{Nature}
  \textbf{\bibinfo{volume}{548}}, \bibinfo{pages}{313} (\bibinfo{year}{2017}).

\bibitem[{\citenamefont{Seo et~al.}(2020)\citenamefont{Seo, Wang, Thomas, Rahn,
  Carmo, Ronning, Bauer, dos Reis, Janoschek, Thompson et~al.}}]{Seo2020}
\bibinfo{author}{\bibfnamefont{S.}~\bibnamefont{Seo}},
  \bibinfo{author}{\bibfnamefont{X.}~\bibnamefont{Wang}},
  \bibinfo{author}{\bibfnamefont{S.~M.} \bibnamefont{Thomas}},
  \bibinfo{author}{\bibfnamefont{M.~C.} \bibnamefont{Rahn}},
  \bibinfo{author}{\bibfnamefont{D.}~\bibnamefont{Carmo}},
  \bibinfo{author}{\bibfnamefont{F.}~\bibnamefont{Ronning}},
  \bibinfo{author}{\bibfnamefont{E.~D.} \bibnamefont{Bauer}},
  \bibinfo{author}{\bibfnamefont{R.~D.} \bibnamefont{dos Reis}},
  \bibinfo{author}{\bibfnamefont{M.}~\bibnamefont{Janoschek}},
  \bibinfo{author}{\bibfnamefont{J.~D.} \bibnamefont{Thompson}},
  \bibnamefont{et~al.}, \bibinfo{journal}{Phys. Rev. X}
  \textbf{\bibinfo{volume}{10}}, \bibinfo{pages}{011035}
  (\bibinfo{year}{2020}).

\bibitem[{\citenamefont{Hecker and Schmalian}(2018)}]{Hecker2018}
\bibinfo{author}{\bibfnamefont{M.}~\bibnamefont{Hecker}} \bibnamefont{and}
  \bibinfo{author}{\bibfnamefont{J.}~\bibnamefont{Schmalian}},
  \bibinfo{journal}{npj Quantum Materials} \textbf{\bibinfo{volume}{3}},
  \bibinfo{pages}{26} (\bibinfo{year}{2018}).

\bibitem[{\citenamefont{Cho et~al.}(2020)\citenamefont{Cho, Shen, Lyu, Atanov,
  Chen, Lee, Hor, Gawryluk, Pomjakushina, Bartkowiak et~al.}}]{Cho2020}
\bibinfo{author}{\bibfnamefont{C.-w.} \bibnamefont{Cho}},
  \bibinfo{author}{\bibfnamefont{J.}~\bibnamefont{Shen}},
  \bibinfo{author}{\bibfnamefont{J.}~\bibnamefont{Lyu}},
  \bibinfo{author}{\bibfnamefont{O.}~\bibnamefont{Atanov}},
  \bibinfo{author}{\bibfnamefont{Q.}~\bibnamefont{Chen}},
  \bibinfo{author}{\bibfnamefont{S.~H.} \bibnamefont{Lee}},
  \bibinfo{author}{\bibfnamefont{Y.~S.} \bibnamefont{Hor}},
  \bibinfo{author}{\bibfnamefont{D.~J.} \bibnamefont{Gawryluk}},
  \bibinfo{author}{\bibfnamefont{E.}~\bibnamefont{Pomjakushina}},
  \bibinfo{author}{\bibfnamefont{M.}~\bibnamefont{Bartkowiak}},
  \bibnamefont{et~al.}, \bibinfo{journal}{Nature Communications}
  \textbf{\bibinfo{volume}{11}}, \bibinfo{pages}{3056} (\bibinfo{year}{2020}),
  ISSN \bibinfo{issn}{2041-1723}.

\bibitem[{\citenamefont{Jin et~al.}(2021)\citenamefont{Jin, Zhang, Guo, Chen,
  Zhou, and Li}}]{cold_atoms_nem}
\bibinfo{author}{\bibfnamefont{S.}~\bibnamefont{Jin}},
  \bibinfo{author}{\bibfnamefont{W.}~\bibnamefont{Zhang}},
  \bibinfo{author}{\bibfnamefont{X.}~\bibnamefont{Guo}},
  \bibinfo{author}{\bibfnamefont{X.}~\bibnamefont{Chen}},
  \bibinfo{author}{\bibfnamefont{X.}~\bibnamefont{Zhou}}, \bibnamefont{and}
  \bibinfo{author}{\bibfnamefont{X.}~\bibnamefont{Li}}, \bibinfo{journal}{Phys.
  Rev. Lett.} \textbf{\bibinfo{volume}{126}}, \bibinfo{pages}{035301}
  (\bibinfo{year}{2021}).

\bibitem[{\citenamefont{Cao et~al.}(2021)\citenamefont{Cao, Rodan-Legrain,
  Park, Yuan, Watanabe, Taniguchi, Fernandes, Fu, and
  Jarillo-Herrero}}]{Cao2021}
\bibinfo{author}{\bibfnamefont{Y.}~\bibnamefont{Cao}},
  \bibinfo{author}{\bibfnamefont{D.}~\bibnamefont{Rodan-Legrain}},
  \bibinfo{author}{\bibfnamefont{J.~M.} \bibnamefont{Park}},
  \bibinfo{author}{\bibfnamefont{N.~F.~Q.} \bibnamefont{Yuan}},
  \bibinfo{author}{\bibfnamefont{K.}~\bibnamefont{Watanabe}},
  \bibinfo{author}{\bibfnamefont{T.}~\bibnamefont{Taniguchi}},
  \bibinfo{author}{\bibfnamefont{R.~M.} \bibnamefont{Fernandes}},
  \bibinfo{author}{\bibfnamefont{L.}~\bibnamefont{Fu}}, \bibnamefont{and}
  \bibinfo{author}{\bibfnamefont{P.}~\bibnamefont{Jarillo-Herrero}},
  \bibinfo{journal}{Science} \textbf{\bibinfo{volume}{372}},
  \bibinfo{pages}{264} (\bibinfo{year}{2021}).

\bibitem[{\citenamefont{Rubio-Verdú et~al.}(2020)\citenamefont{Rubio-Verdú,
  Turkel, Song, Klebl, Samajdar, Scheurer, Venderbos, Watanabe, Taniguchi,
  Ochoa et~al.}}]{Rubio2020}
\bibinfo{author}{\bibfnamefont{C.}~\bibnamefont{Rubio-Verdú}},
  \bibinfo{author}{\bibfnamefont{S.}~\bibnamefont{Turkel}},
  \bibinfo{author}{\bibfnamefont{L.}~\bibnamefont{Song}},
  \bibinfo{author}{\bibfnamefont{L.}~\bibnamefont{Klebl}},
  \bibinfo{author}{\bibfnamefont{R.}~\bibnamefont{Samajdar}},
  \bibinfo{author}{\bibfnamefont{M.~S.} \bibnamefont{Scheurer}},
  \bibinfo{author}{\bibfnamefont{J.~W.~F.} \bibnamefont{Venderbos}},
  \bibinfo{author}{\bibfnamefont{K.}~\bibnamefont{Watanabe}},
  \bibinfo{author}{\bibfnamefont{T.}~\bibnamefont{Taniguchi}},
  \bibinfo{author}{\bibfnamefont{H.}~\bibnamefont{Ochoa}},
  \bibnamefont{et~al.}, \bibinfo{journal}{ArXiv:2009.11645}
  (\bibinfo{year}{2020}).

\bibitem[{\citenamefont{Chu et~al.}(2010)\citenamefont{Chu, Analytis, De~Greve,
  McMahon, Islam, Yamamoto, and Fisher}}]{Chu824}
\bibinfo{author}{\bibfnamefont{J.-H.} \bibnamefont{Chu}},
  \bibinfo{author}{\bibfnamefont{J.~G.} \bibnamefont{Analytis}},
  \bibinfo{author}{\bibfnamefont{K.}~\bibnamefont{De~Greve}},
  \bibinfo{author}{\bibfnamefont{P.~L.} \bibnamefont{McMahon}},
  \bibinfo{author}{\bibfnamefont{Z.}~\bibnamefont{Islam}},
  \bibinfo{author}{\bibfnamefont{Y.}~\bibnamefont{Yamamoto}}, \bibnamefont{and}
  \bibinfo{author}{\bibfnamefont{I.~R.} \bibnamefont{Fisher}},
  \bibinfo{journal}{Science} \textbf{\bibinfo{volume}{329}},
  \bibinfo{pages}{824} (\bibinfo{year}{2010}).

\bibitem[{\citenamefont{Chuang et~al.}(2010)\citenamefont{Chuang, Allan, Lee,
  Xie, Ni, Bud{\textquoteright}ko, Boebinger, Canfield, and Davis}}]{Chuang181}
\bibinfo{author}{\bibfnamefont{T.-M.} \bibnamefont{Chuang}},
  \bibinfo{author}{\bibfnamefont{M.~P.} \bibnamefont{Allan}},
  \bibinfo{author}{\bibfnamefont{J.}~\bibnamefont{Lee}},
  \bibinfo{author}{\bibfnamefont{Y.}~\bibnamefont{Xie}},
  \bibinfo{author}{\bibfnamefont{N.}~\bibnamefont{Ni}},
  \bibinfo{author}{\bibfnamefont{S.~L.} \bibnamefont{Bud{\textquoteright}ko}},
  \bibinfo{author}{\bibfnamefont{G.~S.} \bibnamefont{Boebinger}},
  \bibinfo{author}{\bibfnamefont{P.~C.} \bibnamefont{Canfield}},
  \bibnamefont{and} \bibinfo{author}{\bibfnamefont{J.~C.} \bibnamefont{Davis}},
  \bibinfo{journal}{Science} \textbf{\bibinfo{volume}{327}},
  \bibinfo{pages}{181} (\bibinfo{year}{2010}).

\bibitem[{\citenamefont{Fernandes and Schmalian}(2012)}]{Fernandes_2012}
\bibinfo{author}{\bibfnamefont{R.~M.} \bibnamefont{Fernandes}}
  \bibnamefont{and}
  \bibinfo{author}{\bibfnamefont{J.}~\bibnamefont{Schmalian}},
  \bibinfo{journal}{Superconductor Science and Technology}
  \textbf{\bibinfo{volume}{25}}, \bibinfo{pages}{084005}
  (\bibinfo{year}{2012}).

\bibitem[{\citenamefont{Böhmer and Meingast}(2016)}]{BOHMER201690}
\bibinfo{author}{\bibfnamefont{A.~E.} \bibnamefont{Böhmer}} \bibnamefont{and}
  \bibinfo{author}{\bibfnamefont{C.}~\bibnamefont{Meingast}},
  \bibinfo{journal}{Comptes Rendus Physique} \textbf{\bibinfo{volume}{17}},
  \bibinfo{pages}{90} (\bibinfo{year}{2016}).

\bibitem[{\citenamefont{Gallais and Paul}(2016)}]{GALLAIS2016}
\bibinfo{author}{\bibfnamefont{Y.}~\bibnamefont{Gallais}} \bibnamefont{and}
  \bibinfo{author}{\bibfnamefont{I.}~\bibnamefont{Paul}},
  \bibinfo{journal}{Comptes Rendus Physique} \textbf{\bibinfo{volume}{17}},
  \bibinfo{pages}{113 } (\bibinfo{year}{2016}).

\bibitem[{\citenamefont{Fernandes et~al.}(2014)\citenamefont{Fernandes,
  Chubukov, and Schmalian}}]{Fernandes2014}
\bibinfo{author}{\bibfnamefont{R.~M.} \bibnamefont{Fernandes}},
  \bibinfo{author}{\bibfnamefont{A.~V.} \bibnamefont{Chubukov}},
  \bibnamefont{and}
  \bibinfo{author}{\bibfnamefont{J.}~\bibnamefont{Schmalian}},
  \bibinfo{journal}{Nature Physics} \textbf{\bibinfo{volume}{10}},
  \bibinfo{pages}{97} (\bibinfo{year}{2014}).

\bibitem[{\citenamefont{Kasahara et~al.}(2012)\citenamefont{Kasahara, Shi,
  Hashimoto, Tonegawa, Mizukami, Shibauchi, Sugimoto, Fukuda, Terashima,
  Nevidomskyy et~al.}}]{Kasahara2012}
\bibinfo{author}{\bibfnamefont{S.}~\bibnamefont{Kasahara}},
  \bibinfo{author}{\bibfnamefont{H.~J.} \bibnamefont{Shi}},
  \bibinfo{author}{\bibfnamefont{K.}~\bibnamefont{Hashimoto}},
  \bibinfo{author}{\bibfnamefont{S.}~\bibnamefont{Tonegawa}},
  \bibinfo{author}{\bibfnamefont{Y.}~\bibnamefont{Mizukami}},
  \bibinfo{author}{\bibfnamefont{T.}~\bibnamefont{Shibauchi}},
  \bibinfo{author}{\bibfnamefont{K.}~\bibnamefont{Sugimoto}},
  \bibinfo{author}{\bibfnamefont{T.}~\bibnamefont{Fukuda}},
  \bibinfo{author}{\bibfnamefont{T.}~\bibnamefont{Terashima}},
  \bibinfo{author}{\bibfnamefont{A.~H.} \bibnamefont{Nevidomskyy}},
  \bibnamefont{et~al.}, \bibinfo{journal}{Nature}
  \textbf{\bibinfo{volume}{486}}, \bibinfo{pages}{382} (\bibinfo{year}{2012}).

\bibitem[{\citenamefont{Thewalt et~al.}(2018)\citenamefont{Thewalt, Hayes,
  Hinton, Little, Patankar, Wu, Helm, Stan, Tamura, Analytis
  et~al.}}]{PhysRevLett.121.027001}
\bibinfo{author}{\bibfnamefont{E.}~\bibnamefont{Thewalt}},
  \bibinfo{author}{\bibfnamefont{I.~M.} \bibnamefont{Hayes}},
  \bibinfo{author}{\bibfnamefont{J.~P.} \bibnamefont{Hinton}},
  \bibinfo{author}{\bibfnamefont{A.}~\bibnamefont{Little}},
  \bibinfo{author}{\bibfnamefont{S.}~\bibnamefont{Patankar}},
  \bibinfo{author}{\bibfnamefont{L.}~\bibnamefont{Wu}},
  \bibinfo{author}{\bibfnamefont{T.}~\bibnamefont{Helm}},
  \bibinfo{author}{\bibfnamefont{C.~V.} \bibnamefont{Stan}},
  \bibinfo{author}{\bibfnamefont{N.}~\bibnamefont{Tamura}},
  \bibinfo{author}{\bibfnamefont{J.~G.} \bibnamefont{Analytis}},
  \bibnamefont{et~al.}, \bibinfo{journal}{Phys. Rev. Lett.}
  \textbf{\bibinfo{volume}{121}}, \bibinfo{pages}{027001}
  (\bibinfo{year}{2018}).

\bibitem[{\citenamefont{Iye et~al.}(2015)\citenamefont{Iye, Julien, Mayaffre,
  Horvatić, Berthier, Ishida, Ikeda, Kasahara, Shibauchi, and
  Matsuda}}]{doi:10.7566/JPSJ.84.043705}
\bibinfo{author}{\bibfnamefont{T.}~\bibnamefont{Iye}},
  \bibinfo{author}{\bibfnamefont{M.-H.} \bibnamefont{Julien}},
  \bibinfo{author}{\bibfnamefont{H.}~\bibnamefont{Mayaffre}},
  \bibinfo{author}{\bibfnamefont{M.}~\bibnamefont{Horvatić}},
  \bibinfo{author}{\bibfnamefont{C.}~\bibnamefont{Berthier}},
  \bibinfo{author}{\bibfnamefont{K.}~\bibnamefont{Ishida}},
  \bibinfo{author}{\bibfnamefont{H.}~\bibnamefont{Ikeda}},
  \bibinfo{author}{\bibfnamefont{S.}~\bibnamefont{Kasahara}},
  \bibinfo{author}{\bibfnamefont{T.}~\bibnamefont{Shibauchi}},
  \bibnamefont{and} \bibinfo{author}{\bibfnamefont{Y.}~\bibnamefont{Matsuda}},
  \bibinfo{journal}{Journal of the Physical Society of Japan}
  \textbf{\bibinfo{volume}{84}}, \bibinfo{pages}{043705}
  (\bibinfo{year}{2015}).

\bibitem[{\citenamefont{Rosenthal et~al.}(2014)\citenamefont{Rosenthal,
  Andrade, Arguello, Fernandes, Xing, Wang, Jin, Millis, and
  Pasupathy}}]{Rosenthal2014}
\bibinfo{author}{\bibfnamefont{E.~P.} \bibnamefont{Rosenthal}},
  \bibinfo{author}{\bibfnamefont{E.~F.} \bibnamefont{Andrade}},
  \bibinfo{author}{\bibfnamefont{C.~J.} \bibnamefont{Arguello}},
  \bibinfo{author}{\bibfnamefont{R.~M.} \bibnamefont{Fernandes}},
  \bibinfo{author}{\bibfnamefont{L.~Y.} \bibnamefont{Xing}},
  \bibinfo{author}{\bibfnamefont{X.~C.} \bibnamefont{Wang}},
  \bibinfo{author}{\bibfnamefont{C.~Q.} \bibnamefont{Jin}},
  \bibinfo{author}{\bibfnamefont{A.~J.} \bibnamefont{Millis}},
  \bibnamefont{and} \bibinfo{author}{\bibfnamefont{A.~N.}
  \bibnamefont{Pasupathy}}, \bibinfo{journal}{Nature Physics}
  \textbf{\bibinfo{volume}{10}}, \bibinfo{pages}{225} (\bibinfo{year}{2014}),
  ISSN \bibinfo{issn}{1745-2481}.

\bibitem[{\citenamefont{Man et~al.}(2015)\citenamefont{Man, Lu, Chen, Zhang,
  Zhang, Luo, Kulda, Ivanov, Keller, Morosan et~al.}}]{PhysRevB.92.134521}
\bibinfo{author}{\bibfnamefont{H.}~\bibnamefont{Man}},
  \bibinfo{author}{\bibfnamefont{X.}~\bibnamefont{Lu}},
  \bibinfo{author}{\bibfnamefont{J.~S.} \bibnamefont{Chen}},
  \bibinfo{author}{\bibfnamefont{R.}~\bibnamefont{Zhang}},
  \bibinfo{author}{\bibfnamefont{W.}~\bibnamefont{Zhang}},
  \bibinfo{author}{\bibfnamefont{H.}~\bibnamefont{Luo}},
  \bibinfo{author}{\bibfnamefont{J.}~\bibnamefont{Kulda}},
  \bibinfo{author}{\bibfnamefont{A.}~\bibnamefont{Ivanov}},
  \bibinfo{author}{\bibfnamefont{T.}~\bibnamefont{Keller}},
  \bibinfo{author}{\bibfnamefont{E.}~\bibnamefont{Morosan}},
  \bibnamefont{et~al.}, \bibinfo{journal}{Phys. Rev. B}
  \textbf{\bibinfo{volume}{92}}, \bibinfo{pages}{134521}
  (\bibinfo{year}{2015}).

\bibitem[{\citenamefont{Song and
  Koshelev}(2016{\natexlab{a}})}]{PhysRevB.94.094509}
\bibinfo{author}{\bibfnamefont{K.~W.} \bibnamefont{Song}} \bibnamefont{and}
  \bibinfo{author}{\bibfnamefont{A.~E.} \bibnamefont{Koshelev}},
  \bibinfo{journal}{Phys. Rev. B} \textbf{\bibinfo{volume}{94}},
  \bibinfo{pages}{094509} (\bibinfo{year}{2016}{\natexlab{a}}).

\bibitem[{\citenamefont{Stojchevska et~al.}(2012)\citenamefont{Stojchevska,
  Mertelj, Chu, Fisher, and Mihailovic}}]{PhysRevB.86.024519}
\bibinfo{author}{\bibfnamefont{L.}~\bibnamefont{Stojchevska}},
  \bibinfo{author}{\bibfnamefont{T.}~\bibnamefont{Mertelj}},
  \bibinfo{author}{\bibfnamefont{J.-H.} \bibnamefont{Chu}},
  \bibinfo{author}{\bibfnamefont{I.~R.} \bibnamefont{Fisher}},
  \bibnamefont{and}
  \bibinfo{author}{\bibfnamefont{D.}~\bibnamefont{Mihailovic}},
  \bibinfo{journal}{Phys. Rev. B} \textbf{\bibinfo{volume}{86}},
  \bibinfo{pages}{024519} (\bibinfo{year}{2012}).

\bibitem[{\citenamefont{Shimojima et~al.}(2014)\citenamefont{Shimojima, Sonobe,
  Malaeb, Shinada, Chainani, Shin, Yoshida, Ideta, Fujimori, Kumigashira
  et~al.}}]{PhysRevB.89.045101}
\bibinfo{author}{\bibfnamefont{T.}~\bibnamefont{Shimojima}},
  \bibinfo{author}{\bibfnamefont{T.}~\bibnamefont{Sonobe}},
  \bibinfo{author}{\bibfnamefont{W.}~\bibnamefont{Malaeb}},
  \bibinfo{author}{\bibfnamefont{K.}~\bibnamefont{Shinada}},
  \bibinfo{author}{\bibfnamefont{A.}~\bibnamefont{Chainani}},
  \bibinfo{author}{\bibfnamefont{S.}~\bibnamefont{Shin}},
  \bibinfo{author}{\bibfnamefont{T.}~\bibnamefont{Yoshida}},
  \bibinfo{author}{\bibfnamefont{S.}~\bibnamefont{Ideta}},
  \bibinfo{author}{\bibfnamefont{A.}~\bibnamefont{Fujimori}},
  \bibinfo{author}{\bibfnamefont{H.}~\bibnamefont{Kumigashira}},
  \bibnamefont{et~al.}, \bibinfo{journal}{Phys. Rev. B}
  \textbf{\bibinfo{volume}{89}}, \bibinfo{pages}{045101}
  (\bibinfo{year}{2014}).

\bibitem[{\citenamefont{Sonobe et~al.}(2018)\citenamefont{Sonobe, Shimojima,
  Nakamura, Nakajima, Uchida, Kihou, Lee, Iyo, Eisaki, Ohgushi
  et~al.}}]{Sonobe2018}
\bibinfo{author}{\bibfnamefont{T.}~\bibnamefont{Sonobe}},
  \bibinfo{author}{\bibfnamefont{T.}~\bibnamefont{Shimojima}},
  \bibinfo{author}{\bibfnamefont{A.}~\bibnamefont{Nakamura}},
  \bibinfo{author}{\bibfnamefont{M.}~\bibnamefont{Nakajima}},
  \bibinfo{author}{\bibfnamefont{S.}~\bibnamefont{Uchida}},
  \bibinfo{author}{\bibfnamefont{K.}~\bibnamefont{Kihou}},
  \bibinfo{author}{\bibfnamefont{C.~H.} \bibnamefont{Lee}},
  \bibinfo{author}{\bibfnamefont{A.}~\bibnamefont{Iyo}},
  \bibinfo{author}{\bibfnamefont{H.}~\bibnamefont{Eisaki}},
  \bibinfo{author}{\bibfnamefont{K.}~\bibnamefont{Ohgushi}},
  \bibnamefont{et~al.}, \bibinfo{journal}{Scientific Reports}
  \textbf{\bibinfo{volume}{8}}, \bibinfo{pages}{2169} (\bibinfo{year}{2018}).

\bibitem[{\citenamefont{Zhang et~al.}(2015)\citenamefont{Zhang, Qian, Richard,
  Wang, Miao, Lv, Fu, Wolf, Meingast, Wu et~al.}}]{PhysRevB.91.214503}
\bibinfo{author}{\bibfnamefont{P.}~\bibnamefont{Zhang}},
  \bibinfo{author}{\bibfnamefont{T.}~\bibnamefont{Qian}},
  \bibinfo{author}{\bibfnamefont{P.}~\bibnamefont{Richard}},
  \bibinfo{author}{\bibfnamefont{X.~P.} \bibnamefont{Wang}},
  \bibinfo{author}{\bibfnamefont{H.}~\bibnamefont{Miao}},
  \bibinfo{author}{\bibfnamefont{B.~Q.} \bibnamefont{Lv}},
  \bibinfo{author}{\bibfnamefont{B.~B.} \bibnamefont{Fu}},
  \bibinfo{author}{\bibfnamefont{T.}~\bibnamefont{Wolf}},
  \bibinfo{author}{\bibfnamefont{C.}~\bibnamefont{Meingast}},
  \bibinfo{author}{\bibfnamefont{X.~X.} \bibnamefont{Wu}},
  \bibnamefont{et~al.}, \bibinfo{journal}{Phys. Rev. B}
  \textbf{\bibinfo{volume}{91}}, \bibinfo{pages}{214503}
  (\bibinfo{year}{2015}).

\bibitem[{\citenamefont{Toyoda et~al.}(2018)\citenamefont{Toyoda, Kobayashi,
  and Itoh}}]{PhysRevB.97.094515}
\bibinfo{author}{\bibfnamefont{M.}~\bibnamefont{Toyoda}},
  \bibinfo{author}{\bibfnamefont{Y.}~\bibnamefont{Kobayashi}},
  \bibnamefont{and} \bibinfo{author}{\bibfnamefont{M.}~\bibnamefont{Itoh}},
  \bibinfo{journal}{Phys. Rev. B} \textbf{\bibinfo{volume}{97}},
  \bibinfo{pages}{094515} (\bibinfo{year}{2018}).

\bibitem[{\citenamefont{Wiecki et~al.}(2017)\citenamefont{Wiecki, Nandi,
  B\"ohmer, Bud'ko, Canfield, and Furukawa}}]{PhysRevB.96.180502}
\bibinfo{author}{\bibfnamefont{P.}~\bibnamefont{Wiecki}},
  \bibinfo{author}{\bibfnamefont{M.}~\bibnamefont{Nandi}},
  \bibinfo{author}{\bibfnamefont{A.~E.} \bibnamefont{B\"ohmer}},
  \bibinfo{author}{\bibfnamefont{S.~L.} \bibnamefont{Bud'ko}},
  \bibinfo{author}{\bibfnamefont{P.~C.} \bibnamefont{Canfield}},
  \bibnamefont{and} \bibinfo{author}{\bibfnamefont{Y.}~\bibnamefont{Furukawa}},
  \bibinfo{journal}{Phys. Rev. B} \textbf{\bibinfo{volume}{96}},
  \bibinfo{pages}{180502} (\bibinfo{year}{2017}).

\bibitem[{\citenamefont{Yuan et~al.}(2021)\citenamefont{Yuan, Fan, Wang, He,
  Zhang, Xue, and Li}}]{Yuan2021}
\bibinfo{author}{\bibfnamefont{Y.}~\bibnamefont{Yuan}},
  \bibinfo{author}{\bibfnamefont{X.}~\bibnamefont{Fan}},
  \bibinfo{author}{\bibfnamefont{X.}~\bibnamefont{Wang}},
  \bibinfo{author}{\bibfnamefont{K.}~\bibnamefont{He}},
  \bibinfo{author}{\bibfnamefont{Y.}~\bibnamefont{Zhang}},
  \bibinfo{author}{\bibfnamefont{Q.-K.} \bibnamefont{Xue}}, \bibnamefont{and}
  \bibinfo{author}{\bibfnamefont{W.}~\bibnamefont{Li}},
  \bibinfo{journal}{Nature Communications} \textbf{\bibinfo{volume}{12}},
  \bibinfo{pages}{2196} (\bibinfo{year}{2021}).

\bibitem[{\citenamefont{Li et~al.}(2017)\citenamefont{Li, Zhang, Deng, Xu, Mo,
  Yi, Ding, Hashimoto, Moore, Lu et~al.}}]{Li2017}
\bibinfo{author}{\bibfnamefont{W.}~\bibnamefont{Li}},
  \bibinfo{author}{\bibfnamefont{Y.}~\bibnamefont{Zhang}},
  \bibinfo{author}{\bibfnamefont{P.}~\bibnamefont{Deng}},
  \bibinfo{author}{\bibfnamefont{Z.}~\bibnamefont{Xu}},
  \bibinfo{author}{\bibfnamefont{S.-K.} \bibnamefont{Mo}},
  \bibinfo{author}{\bibfnamefont{M.}~\bibnamefont{Yi}},
  \bibinfo{author}{\bibfnamefont{H.}~\bibnamefont{Ding}},
  \bibinfo{author}{\bibfnamefont{M.}~\bibnamefont{Hashimoto}},
  \bibinfo{author}{\bibfnamefont{R.~G.} \bibnamefont{Moore}},
  \bibinfo{author}{\bibfnamefont{D.-H.} \bibnamefont{Lu}},
  \bibnamefont{et~al.}, \bibinfo{journal}{Nature Physics}
  \textbf{\bibinfo{volume}{13}}, \bibinfo{pages}{957} (\bibinfo{year}{2017}).

\bibitem[{\citenamefont{Yim et~al.}(2018)\citenamefont{Yim, Trainer, Aluru,
  Chi, Hardy, Liang, Bonn, and Wahl}}]{Yim2018}
\bibinfo{author}{\bibfnamefont{C.~M.} \bibnamefont{Yim}},
  \bibinfo{author}{\bibfnamefont{C.}~\bibnamefont{Trainer}},
  \bibinfo{author}{\bibfnamefont{R.}~\bibnamefont{Aluru}},
  \bibinfo{author}{\bibfnamefont{S.}~\bibnamefont{Chi}},
  \bibinfo{author}{\bibfnamefont{W.~N.} \bibnamefont{Hardy}},
  \bibinfo{author}{\bibfnamefont{R.}~\bibnamefont{Liang}},
  \bibinfo{author}{\bibfnamefont{D.}~\bibnamefont{Bonn}}, \bibnamefont{and}
  \bibinfo{author}{\bibfnamefont{P.}~\bibnamefont{Wahl}},
  \bibinfo{journal}{Nature Communications} \textbf{\bibinfo{volume}{9}},
  \bibinfo{pages}{2602} (\bibinfo{year}{2018}).

\bibitem[{\citenamefont{Shimojima et~al.}(2021)\citenamefont{Shimojima,
  Motoyui, Taniuchi, Bareille, Onari, Kontani, Nakajima, Kasahara, Shibauchi,
  Matsuda et~al.}}]{Shimojima2021}
\bibinfo{author}{\bibfnamefont{T.}~\bibnamefont{Shimojima}},
  \bibinfo{author}{\bibfnamefont{Y.}~\bibnamefont{Motoyui}},
  \bibinfo{author}{\bibfnamefont{T.}~\bibnamefont{Taniuchi}},
  \bibinfo{author}{\bibfnamefont{C.}~\bibnamefont{Bareille}},
  \bibinfo{author}{\bibfnamefont{S.}~\bibnamefont{Onari}},
  \bibinfo{author}{\bibfnamefont{H.}~\bibnamefont{Kontani}},
  \bibinfo{author}{\bibfnamefont{M.}~\bibnamefont{Nakajima}},
  \bibinfo{author}{\bibfnamefont{S.}~\bibnamefont{Kasahara}},
  \bibinfo{author}{\bibfnamefont{T.}~\bibnamefont{Shibauchi}},
  \bibinfo{author}{\bibfnamefont{Y.}~\bibnamefont{Matsuda}},
  \bibnamefont{et~al.}, \bibinfo{journal}{Science}
  \textbf{\bibinfo{volume}{373}}, \bibinfo{pages}{1122} (\bibinfo{year}{2021}).

\bibitem[{\citenamefont{Luo et~al.}(2015)\citenamefont{Luo, Stanev, Shen, Fang,
  Ling, Osborn, Rosenkranz, Benseman, Divan, Kwok et~al.}}]{PhysRevB.91.094512}
\bibinfo{author}{\bibfnamefont{X.}~\bibnamefont{Luo}},
  \bibinfo{author}{\bibfnamefont{V.}~\bibnamefont{Stanev}},
  \bibinfo{author}{\bibfnamefont{B.}~\bibnamefont{Shen}},
  \bibinfo{author}{\bibfnamefont{L.}~\bibnamefont{Fang}},
  \bibinfo{author}{\bibfnamefont{X.~S.} \bibnamefont{Ling}},
  \bibinfo{author}{\bibfnamefont{R.}~\bibnamefont{Osborn}},
  \bibinfo{author}{\bibfnamefont{S.}~\bibnamefont{Rosenkranz}},
  \bibinfo{author}{\bibfnamefont{T.~M.} \bibnamefont{Benseman}},
  \bibinfo{author}{\bibfnamefont{R.}~\bibnamefont{Divan}},
  \bibinfo{author}{\bibfnamefont{W.-K.} \bibnamefont{Kwok}},
  \bibnamefont{et~al.}, \bibinfo{journal}{Phys. Rev. B}
  \textbf{\bibinfo{volume}{91}}, \bibinfo{pages}{094512}
  (\bibinfo{year}{2015}).

\bibitem[{\citenamefont{Cardy}(1996)}]{Cardy1996}
\bibinfo{author}{\bibfnamefont{J.}~\bibnamefont{Cardy}},
  \emph{\bibinfo{title}{Scaling and Renormalization in Statistical Physics}}
  (\bibinfo{publisher}{Cambridge University Press}, \bibinfo{year}{1996}).

\bibitem[{\citenamefont{Song and Koshelev}(2016{\natexlab{b}})}]{Koshelev2016}
\bibinfo{author}{\bibfnamefont{K.~W.} \bibnamefont{Song}} \bibnamefont{and}
  \bibinfo{author}{\bibfnamefont{A.~E.} \bibnamefont{Koshelev}},
  \bibinfo{journal}{Phys. Rev. B} \textbf{\bibinfo{volume}{94}},
  \bibinfo{pages}{094509} (\bibinfo{year}{2016}{\natexlab{b}}).

\bibitem[{\citenamefont{Fernandes et~al.}(2010)\citenamefont{Fernandes,
  VanBebber, Bhattacharya, Chandra, Keppens, Mandrus, McGuire, Sales, Sefat,
  and Schmalian}}]{PhysRevLett.105.157003}
\bibinfo{author}{\bibfnamefont{R.~M.} \bibnamefont{Fernandes}},
  \bibinfo{author}{\bibfnamefont{L.~H.} \bibnamefont{VanBebber}},
  \bibinfo{author}{\bibfnamefont{S.}~\bibnamefont{Bhattacharya}},
  \bibinfo{author}{\bibfnamefont{P.}~\bibnamefont{Chandra}},
  \bibinfo{author}{\bibfnamefont{V.}~\bibnamefont{Keppens}},
  \bibinfo{author}{\bibfnamefont{D.}~\bibnamefont{Mandrus}},
  \bibinfo{author}{\bibfnamefont{M.~A.} \bibnamefont{McGuire}},
  \bibinfo{author}{\bibfnamefont{B.~C.} \bibnamefont{Sales}},
  \bibinfo{author}{\bibfnamefont{A.~S.} \bibnamefont{Sefat}}, \bibnamefont{and}
  \bibinfo{author}{\bibfnamefont{J.}~\bibnamefont{Schmalian}},
  \bibinfo{journal}{Phys. Rev. Lett.} \textbf{\bibinfo{volume}{105}},
  \bibinfo{pages}{157003} (\bibinfo{year}{2010}).

\bibitem[{\citenamefont{Goto et~al.}(2011)\citenamefont{Goto, Kurihara, Araki,
  Mitsumoto, Akatsu, Nemoto, Tatematsu, and Sato}}]{doi:10.1143/JPSJ.80.073702}
\bibinfo{author}{\bibfnamefont{T.}~\bibnamefont{Goto}},
  \bibinfo{author}{\bibfnamefont{R.}~\bibnamefont{Kurihara}},
  \bibinfo{author}{\bibfnamefont{K.}~\bibnamefont{Araki}},
  \bibinfo{author}{\bibfnamefont{K.}~\bibnamefont{Mitsumoto}},
  \bibinfo{author}{\bibfnamefont{M.}~\bibnamefont{Akatsu}},
  \bibinfo{author}{\bibfnamefont{Y.}~\bibnamefont{Nemoto}},
  \bibinfo{author}{\bibfnamefont{S.}~\bibnamefont{Tatematsu}},
  \bibnamefont{and} \bibinfo{author}{\bibfnamefont{M.}~\bibnamefont{Sato}},
  \bibinfo{journal}{Journal of the Physical Society of Japan}
  \textbf{\bibinfo{volume}{80}}, \bibinfo{pages}{073702}
  (\bibinfo{year}{2011}).

\bibitem[{\citenamefont{Yoshizawa et~al.}(2012)\citenamefont{Yoshizawa, Kimura,
  Chiba, Simayi, Nakanishi, Kihou, Lee, Iyo, Eisaki, Nakajima
  et~al.}}]{doi:10.1143/JPSJ.81.024604}
\bibinfo{author}{\bibfnamefont{M.}~\bibnamefont{Yoshizawa}},
  \bibinfo{author}{\bibfnamefont{D.}~\bibnamefont{Kimura}},
  \bibinfo{author}{\bibfnamefont{T.}~\bibnamefont{Chiba}},
  \bibinfo{author}{\bibfnamefont{S.}~\bibnamefont{Simayi}},
  \bibinfo{author}{\bibfnamefont{Y.}~\bibnamefont{Nakanishi}},
  \bibinfo{author}{\bibfnamefont{K.}~\bibnamefont{Kihou}},
  \bibinfo{author}{\bibfnamefont{C.-H.} \bibnamefont{Lee}},
  \bibinfo{author}{\bibfnamefont{A.}~\bibnamefont{Iyo}},
  \bibinfo{author}{\bibfnamefont{H.}~\bibnamefont{Eisaki}},
  \bibinfo{author}{\bibfnamefont{M.}~\bibnamefont{Nakajima}},
  \bibnamefont{et~al.}, \bibinfo{journal}{Journal of the Physical Society of
  Japan} \textbf{\bibinfo{volume}{81}}, \bibinfo{pages}{024604}
  (\bibinfo{year}{2012}).

\bibitem[{\citenamefont{Fernandes et~al.}(2013)\citenamefont{Fernandes,
  B\"ohmer, Meingast, and Schmalian}}]{PhysRevLett.111.137001}
\bibinfo{author}{\bibfnamefont{R.~M.} \bibnamefont{Fernandes}},
  \bibinfo{author}{\bibfnamefont{A.~E.} \bibnamefont{B\"ohmer}},
  \bibinfo{author}{\bibfnamefont{C.}~\bibnamefont{Meingast}}, \bibnamefont{and}
  \bibinfo{author}{\bibfnamefont{J.}~\bibnamefont{Schmalian}},
  \bibinfo{journal}{Phys. Rev. Lett.} \textbf{\bibinfo{volume}{111}},
  \bibinfo{pages}{137001} (\bibinfo{year}{2013}).

\bibitem[{\citenamefont{Merritt et~al.}(2020)\citenamefont{Merritt, Weber,
  Castellan, Wolf, Ishikawa, Said, Alatas, Fernandes, Baron, and
  Reznik}}]{PhysRevLett.124.157001}
\bibinfo{author}{\bibfnamefont{A.~M.} \bibnamefont{Merritt}},
  \bibinfo{author}{\bibfnamefont{F.}~\bibnamefont{Weber}},
  \bibinfo{author}{\bibfnamefont{J.-P.} \bibnamefont{Castellan}},
  \bibinfo{author}{\bibfnamefont{T.}~\bibnamefont{Wolf}},
  \bibinfo{author}{\bibfnamefont{D.}~\bibnamefont{Ishikawa}},
  \bibinfo{author}{\bibfnamefont{A.~H.} \bibnamefont{Said}},
  \bibinfo{author}{\bibfnamefont{A.}~\bibnamefont{Alatas}},
  \bibinfo{author}{\bibfnamefont{R.~M.} \bibnamefont{Fernandes}},
  \bibinfo{author}{\bibfnamefont{A.~Q.~R.} \bibnamefont{Baron}},
  \bibnamefont{and} \bibinfo{author}{\bibfnamefont{D.}~\bibnamefont{Reznik}},
  \bibinfo{journal}{Phys. Rev. Lett.} \textbf{\bibinfo{volume}{124}},
  \bibinfo{pages}{157001} (\bibinfo{year}{2020}).

\bibitem[{\citenamefont{Chibani et~al.}(2021)\citenamefont{Chibani, Farina,
  Massat, Cazayous, Sacuto, Urata, Tanabe, Tanigaki, B{\"o}hmer, Canfield
  et~al.}}]{Chibani2021}
\bibinfo{author}{\bibfnamefont{S.}~\bibnamefont{Chibani}},
  \bibinfo{author}{\bibfnamefont{D.}~\bibnamefont{Farina}},
  \bibinfo{author}{\bibfnamefont{P.}~\bibnamefont{Massat}},
  \bibinfo{author}{\bibfnamefont{M.}~\bibnamefont{Cazayous}},
  \bibinfo{author}{\bibfnamefont{A.}~\bibnamefont{Sacuto}},
  \bibinfo{author}{\bibfnamefont{T.}~\bibnamefont{Urata}},
  \bibinfo{author}{\bibfnamefont{Y.}~\bibnamefont{Tanabe}},
  \bibinfo{author}{\bibfnamefont{K.}~\bibnamefont{Tanigaki}},
  \bibinfo{author}{\bibfnamefont{A.~E.} \bibnamefont{B{\"o}hmer}},
  \bibinfo{author}{\bibfnamefont{P.~C.} \bibnamefont{Canfield}},
  \bibnamefont{et~al.}, \bibinfo{journal}{npj Quantum Materials}
  \textbf{\bibinfo{volume}{6}}, \bibinfo{pages}{37} (\bibinfo{year}{2021}).

\bibitem[{\citenamefont{Qi and Xu}(2009)}]{Qi2009}
\bibinfo{author}{\bibfnamefont{Y.}~\bibnamefont{Qi}} \bibnamefont{and}
  \bibinfo{author}{\bibfnamefont{C.}~\bibnamefont{Xu}}, \bibinfo{journal}{Phys.
  Rev. B} \textbf{\bibinfo{volume}{80}}, \bibinfo{pages}{094402}
  (\bibinfo{year}{2009}).

\bibitem[{\citenamefont{Karahasanovic and Schmalian}(2016)}]{Schmalian2016}
\bibinfo{author}{\bibfnamefont{U.}~\bibnamefont{Karahasanovic}}
  \bibnamefont{and}
  \bibinfo{author}{\bibfnamefont{J.}~\bibnamefont{Schmalian}},
  \bibinfo{journal}{Phys. Rev. B} \textbf{\bibinfo{volume}{93}},
  \bibinfo{pages}{064520} (\bibinfo{year}{2016}).

\bibitem[{\citenamefont{Paul and Garst}(2017)}]{Paul_Garst}
\bibinfo{author}{\bibfnamefont{I.}~\bibnamefont{Paul}} \bibnamefont{and}
  \bibinfo{author}{\bibfnamefont{M.}~\bibnamefont{Garst}},
  \bibinfo{journal}{Phys. Rev. Lett.} \textbf{\bibinfo{volume}{118}},
  \bibinfo{pages}{227601} (\bibinfo{year}{2017}).

\bibitem[{\citenamefont{de~Carvalho and Fernandes}(2019)}]{Carvalho2019}
\bibinfo{author}{\bibfnamefont{V.~S.} \bibnamefont{de~Carvalho}}
  \bibnamefont{and} \bibinfo{author}{\bibfnamefont{R.~M.}
  \bibnamefont{Fernandes}}, \bibinfo{journal}{Phys. Rev. B}
  \textbf{\bibinfo{volume}{100}}, \bibinfo{pages}{115103}
  (\bibinfo{year}{2019}).

\bibitem[{\citenamefont{Carlson et~al.}(2006)\citenamefont{Carlson, Dahmen,
  Fradkin, and Kivelson}}]{Carlson2006}
\bibinfo{author}{\bibfnamefont{E.~W.} \bibnamefont{Carlson}},
  \bibinfo{author}{\bibfnamefont{K.~A.} \bibnamefont{Dahmen}},
  \bibinfo{author}{\bibfnamefont{E.}~\bibnamefont{Fradkin}}, \bibnamefont{and}
  \bibinfo{author}{\bibfnamefont{S.~A.} \bibnamefont{Kivelson}},
  \bibinfo{journal}{Phys. Rev. Lett.} \textbf{\bibinfo{volume}{96}},
  \bibinfo{pages}{097003} (\bibinfo{year}{2006}).

\bibitem[{\citenamefont{Kuo et~al.}(2016)\citenamefont{Kuo, Chu, Palmstrom,
  Kivelson, and Fisher}}]{Kuo2016}
\bibinfo{author}{\bibfnamefont{H.-H.} \bibnamefont{Kuo}},
  \bibinfo{author}{\bibfnamefont{J.-H.} \bibnamefont{Chu}},
  \bibinfo{author}{\bibfnamefont{J.~C.} \bibnamefont{Palmstrom}},
  \bibinfo{author}{\bibfnamefont{S.~A.} \bibnamefont{Kivelson}},
  \bibnamefont{and} \bibinfo{author}{\bibfnamefont{I.~R.}
  \bibnamefont{Fisher}}, \bibinfo{journal}{Science}
  \textbf{\bibinfo{volume}{352}}, \bibinfo{pages}{958} (\bibinfo{year}{2016}).

\bibitem[{\citenamefont{Wiecki et~al.}(2021)\citenamefont{Wiecki, Zhou, Julien,
  B\"ohmer, and Schmalian}}]{Wiecki2021}
\bibinfo{author}{\bibfnamefont{P.}~\bibnamefont{Wiecki}},
  \bibinfo{author}{\bibfnamefont{R.}~\bibnamefont{Zhou}},
  \bibinfo{author}{\bibfnamefont{M.-H.} \bibnamefont{Julien}},
  \bibinfo{author}{\bibfnamefont{A.~E.} \bibnamefont{B\"ohmer}},
  \bibnamefont{and}
  \bibinfo{author}{\bibfnamefont{J.}~\bibnamefont{Schmalian}},
  \bibinfo{journal}{Phys. Rev. B} \textbf{\bibinfo{volume}{104}},
  \bibinfo{pages}{125134} (\bibinfo{year}{2021}).

\bibitem[{\citenamefont{Lahiri}(2021)}]{athesis}
\bibinfo{author}{\bibfnamefont{A.}~\bibnamefont{Lahiri}}, Master's thesis,
  \bibinfo{school}{University of Minnesota} (\bibinfo{year}{2021}).

\bibitem[{\citenamefont{Lifshitz et~al.}(1986)\citenamefont{Lifshitz, Kosevich,
  and Pitaevskii}}]{LIFSHITZ198687}
\bibinfo{author}{\bibfnamefont{E.}~\bibnamefont{Lifshitz}},
  \bibinfo{author}{\bibfnamefont{A.}~\bibnamefont{Kosevich}}, \bibnamefont{and}
  \bibinfo{author}{\bibfnamefont{L.}~\bibnamefont{Pitaevskii}},
  \emph{\bibinfo{title}{Theory of Elasticity}}
  (\bibinfo{publisher}{Butterworth-Heinemann}, \bibinfo{year}{1986}).

\bibitem[{\citenamefont{Nie et~al.}(2014)\citenamefont{Nie, Tarjus, and
  Kivelson}}]{Nie2014}
\bibinfo{author}{\bibfnamefont{L.}~\bibnamefont{Nie}},
  \bibinfo{author}{\bibfnamefont{G.}~\bibnamefont{Tarjus}}, \bibnamefont{and}
  \bibinfo{author}{\bibfnamefont{S.~A.} \bibnamefont{Kivelson}},
  \bibinfo{journal}{Proceedings of the National Academy of Sciences}
  \textbf{\bibinfo{volume}{111}}, \bibinfo{pages}{7980} (\bibinfo{year}{2014}).

\bibitem[{\citenamefont{Cui and Fernandes}(2018)}]{Cui2018}
\bibinfo{author}{\bibfnamefont{T.}~\bibnamefont{Cui}} \bibnamefont{and}
  \bibinfo{author}{\bibfnamefont{R.~M.} \bibnamefont{Fernandes}},
  \bibinfo{journal}{Phys. Rev. B} \textbf{\bibinfo{volume}{98}},
  \bibinfo{pages}{085117} (\bibinfo{year}{2018}).

\bibitem[{\citenamefont{Marchenko and Parshin}(1980)}]{MarchenkoParshin}
\bibinfo{author}{\bibfnamefont{V.}~\bibnamefont{Marchenko}} \bibnamefont{and}
  \bibinfo{author}{\bibfnamefont{A.~Y.} \bibnamefont{Parshin}},
  \bibinfo{journal}{JETP Lett.} \textbf{\bibinfo{volume}{52}},
  \bibinfo{pages}{129} (\bibinfo{year}{1980}), \bibinfo{note}{[ZhETF, Vol. 79,
  No. 1, p. 257, July 1980]}.

\bibitem[{\citenamefont{Shilkrot and Srolovitz}(1996)}]{PhysRevB.53.11120}
\bibinfo{author}{\bibfnamefont{L.~E.} \bibnamefont{Shilkrot}} \bibnamefont{and}
  \bibinfo{author}{\bibfnamefont{D.~J.} \bibnamefont{Srolovitz}},
  \bibinfo{journal}{Phys. Rev. B} \textbf{\bibinfo{volume}{53}},
  \bibinfo{pages}{11120} (\bibinfo{year}{1996}).

\bibitem[{\citenamefont{Stewart et~al.}(1994)\citenamefont{Stewart, Pohland,
  and Gibson}}]{PhysRevB.49.13848}
\bibinfo{author}{\bibfnamefont{J.}~\bibnamefont{Stewart}},
  \bibinfo{author}{\bibfnamefont{O.}~\bibnamefont{Pohland}}, \bibnamefont{and}
  \bibinfo{author}{\bibfnamefont{J.~M.} \bibnamefont{Gibson}},
  \bibinfo{journal}{Phys. Rev. B} \textbf{\bibinfo{volume}{49}},
  \bibinfo{pages}{13848} (\bibinfo{year}{1994}).

\bibitem[{doi(2005)}]{doi:https://doi.org/10.1002/352760667X.ch9}
\emph{\bibinfo{title}{Point Defects}} (\bibinfo{publisher}{John Wiley {\&}
  Sons, Ltd}, \bibinfo{year}{2005}), ISBN \bibinfo{isbn}{9783527606672}.

\bibitem[{\citenamefont{Bacon et~al.}(1980)\citenamefont{Bacon, Barnett, and
  Scattergood}}]{BACON198051}
\bibinfo{author}{\bibfnamefont{D.}~\bibnamefont{Bacon}},
  \bibinfo{author}{\bibfnamefont{D.}~\bibnamefont{Barnett}}, \bibnamefont{and}
  \bibinfo{author}{\bibfnamefont{R.}~\bibnamefont{Scattergood}},
  \bibinfo{journal}{Progress in Materials Science}
  \textbf{\bibinfo{volume}{23}}, \bibinfo{pages}{51} (\bibinfo{year}{1980}).

\bibitem[{\citenamefont{Clouet et~al.}(2018)\citenamefont{Clouet, Varvenne, and
  Jourdan}}]{CLOUET201849}
\bibinfo{author}{\bibfnamefont{E.}~\bibnamefont{Clouet}},
  \bibinfo{author}{\bibfnamefont{C.}~\bibnamefont{Varvenne}}, \bibnamefont{and}
  \bibinfo{author}{\bibfnamefont{T.}~\bibnamefont{Jourdan}},
  \bibinfo{journal}{Computational Materials Science}
  \textbf{\bibinfo{volume}{147}}, \bibinfo{pages}{49} (\bibinfo{year}{2018}).

\bibitem[{\citenamefont{Teodosiu}(1982)}]{Teodosiu1982}
\bibinfo{author}{\bibfnamefont{C.}~\bibnamefont{Teodosiu}},
  \emph{\bibinfo{title}{Elastic Models of Crystal Defects}}
  (\bibinfo{publisher}{Springer-Verlag}, \bibinfo{year}{1982}),
  \bibinfo{edition}{1st} ed.

\bibitem[{\citenamefont{De~Dominicis and
  Giardina}(2006)}]{dedominicis_giardina_2006}
\bibinfo{author}{\bibfnamefont{C.}~\bibnamefont{De~Dominicis}}
  \bibnamefont{and} \bibinfo{author}{\bibfnamefont{I.}~\bibnamefont{Giardina}},
  \emph{\bibinfo{title}{Random Fields and Spin Glasses: A Field Theory
  Approach}} (\bibinfo{publisher}{Cambridge University Press},
  \bibinfo{year}{2006}).

\bibitem[{\citenamefont{Dotsenko}(2000)}]{dotsenko_2000}
\bibinfo{author}{\bibfnamefont{V.}~\bibnamefont{Dotsenko}},
  \emph{\bibinfo{title}{Introduction to the Replica Theory of Disordered
  Statistical Systems}}, Collection Alea-Saclay: Monographs and Texts in
  Statistical Physics (\bibinfo{publisher}{Cambridge University Press},
  \bibinfo{year}{2000}).

\bibitem[{\citenamefont{Imry and Ma}(1975)}]{PhysRevLett.35.1399}
\bibinfo{author}{\bibfnamefont{Y.}~\bibnamefont{Imry}} \bibnamefont{and}
  \bibinfo{author}{\bibfnamefont{S.-k.} \bibnamefont{Ma}},
  \bibinfo{journal}{Phys. Rev. Lett.} \textbf{\bibinfo{volume}{35}},
  \bibinfo{pages}{1399} (\bibinfo{year}{1975}).

\bibitem[{\citenamefont{Aharony et~al.}(1976)\citenamefont{Aharony, Imry, and
  Ma}}]{PhysRevLett.37.1364}
\bibinfo{author}{\bibfnamefont{A.}~\bibnamefont{Aharony}},
  \bibinfo{author}{\bibfnamefont{Y.}~\bibnamefont{Imry}}, \bibnamefont{and}
  \bibinfo{author}{\bibfnamefont{S.-k.} \bibnamefont{Ma}},
  \bibinfo{journal}{Phys. Rev. Lett.} \textbf{\bibinfo{volume}{37}},
  \bibinfo{pages}{1364} (\bibinfo{year}{1976}).

\bibitem[{\citenamefont{Grinstein}(1976)}]{PhysRevLett.37.944}
\bibinfo{author}{\bibfnamefont{G.}~\bibnamefont{Grinstein}},
  \bibinfo{journal}{Phys. Rev. Lett.} \textbf{\bibinfo{volume}{37}},
  \bibinfo{pages}{944} (\bibinfo{year}{1976}).

\bibitem[{\citenamefont{M{\'{e}}zard and Young}(1992)}]{M_zard_1992}
\bibinfo{author}{\bibfnamefont{M.}~\bibnamefont{M{\'{e}}zard}}
  \bibnamefont{and} \bibinfo{author}{\bibfnamefont{A.~P.} \bibnamefont{Young}},
  \bibinfo{journal}{Europhysics Letters ({EPL})} \textbf{\bibinfo{volume}{18}},
  \bibinfo{pages}{653} (\bibinfo{year}{1992}).

\bibitem[{\citenamefont{Young}(1977)}]{Young_1977}
\bibinfo{author}{\bibfnamefont{A.~P.} \bibnamefont{Young}},
  \bibinfo{journal}{Journal of Physics C: Solid State Physics}
  \textbf{\bibinfo{volume}{10}}, \bibinfo{pages}{L257} (\bibinfo{year}{1977}).

\bibitem[{\citenamefont{Hameed et~al.}(2021)\citenamefont{Hameed, Pelc,
  Anderson, Klein, Spieker, Yue, Das, Ramberger, Lukas, Liu
  et~al.}}]{Hameed2021}
\bibinfo{author}{\bibfnamefont{S.}~\bibnamefont{Hameed}},
  \bibinfo{author}{\bibfnamefont{D.}~\bibnamefont{Pelc}},
  \bibinfo{author}{\bibfnamefont{Z.~W.} \bibnamefont{Anderson}},
  \bibinfo{author}{\bibfnamefont{A.}~\bibnamefont{Klein}},
  \bibinfo{author}{\bibfnamefont{R.~J.} \bibnamefont{Spieker}},
  \bibinfo{author}{\bibfnamefont{L.}~\bibnamefont{Yue}},
  \bibinfo{author}{\bibfnamefont{B.}~\bibnamefont{Das}},
  \bibinfo{author}{\bibfnamefont{J.}~\bibnamefont{Ramberger}},
  \bibinfo{author}{\bibfnamefont{M.}~\bibnamefont{Lukas}},
  \bibinfo{author}{\bibfnamefont{Y.}~\bibnamefont{Liu}}, \bibnamefont{et~al.},
  \bibinfo{journal}{Nature Materials}  (\bibinfo{year}{2021}).

\bibitem[{\citenamefont{Willa et~al.}(2021)\citenamefont{Willa, Hecker,
  Fernandes, and Schmalian}}]{PhysRevB.104.024511}
\bibinfo{author}{\bibfnamefont{R.}~\bibnamefont{Willa}},
  \bibinfo{author}{\bibfnamefont{M.}~\bibnamefont{Hecker}},
  \bibinfo{author}{\bibfnamefont{R.~M.} \bibnamefont{Fernandes}},
  \bibnamefont{and}
  \bibinfo{author}{\bibfnamefont{J.}~\bibnamefont{Schmalian}},
  \bibinfo{journal}{Phys. Rev. B} \textbf{\bibinfo{volume}{104}},
  \bibinfo{pages}{024511} (\bibinfo{year}{2021}).

\bibitem[{\citenamefont{Avci et~al.}(2012)\citenamefont{Avci, Chmaissem, Chung,
  Rosenkranz, Goremychkin, Castellan, Todorov, Schlueter, Claus, Daoud-Aladine
  et~al.}}]{Avci2012}
\bibinfo{author}{\bibfnamefont{S.}~\bibnamefont{Avci}},
  \bibinfo{author}{\bibfnamefont{O.}~\bibnamefont{Chmaissem}},
  \bibinfo{author}{\bibfnamefont{D.~Y.} \bibnamefont{Chung}},
  \bibinfo{author}{\bibfnamefont{S.}~\bibnamefont{Rosenkranz}},
  \bibinfo{author}{\bibfnamefont{E.~A.} \bibnamefont{Goremychkin}},
  \bibinfo{author}{\bibfnamefont{J.~P.} \bibnamefont{Castellan}},
  \bibinfo{author}{\bibfnamefont{I.~S.} \bibnamefont{Todorov}},
  \bibinfo{author}{\bibfnamefont{J.~A.} \bibnamefont{Schlueter}},
  \bibinfo{author}{\bibfnamefont{H.}~\bibnamefont{Claus}},
  \bibinfo{author}{\bibfnamefont{A.}~\bibnamefont{Daoud-Aladine}},
  \bibnamefont{et~al.}, \bibinfo{journal}{Phys. Rev. B}
  \textbf{\bibinfo{volume}{85}}, \bibinfo{pages}{184507}
  (\bibinfo{year}{2012}).

\bibitem[{\citenamefont{Murayama et~al.}(2019)\citenamefont{Murayama, Sato,
  Kurihara, Kasahara, Mizukami, Kasahara, Uchiyama, Yamamoto, Moon, Cai
  et~al.}}]{Murayama2019}
\bibinfo{author}{\bibfnamefont{H.}~\bibnamefont{Murayama}},
  \bibinfo{author}{\bibfnamefont{Y.}~\bibnamefont{Sato}},
  \bibinfo{author}{\bibfnamefont{R.}~\bibnamefont{Kurihara}},
  \bibinfo{author}{\bibfnamefont{S.}~\bibnamefont{Kasahara}},
  \bibinfo{author}{\bibfnamefont{Y.}~\bibnamefont{Mizukami}},
  \bibinfo{author}{\bibfnamefont{Y.}~\bibnamefont{Kasahara}},
  \bibinfo{author}{\bibfnamefont{H.}~\bibnamefont{Uchiyama}},
  \bibinfo{author}{\bibfnamefont{A.}~\bibnamefont{Yamamoto}},
  \bibinfo{author}{\bibfnamefont{E.-G.} \bibnamefont{Moon}},
  \bibinfo{author}{\bibfnamefont{J.}~\bibnamefont{Cai}}, \bibnamefont{et~al.},
  \bibinfo{journal}{Nature Communications} \textbf{\bibinfo{volume}{10}},
  \bibinfo{pages}{3282} (\bibinfo{year}{2019}).

\bibitem[{\citenamefont{Shimojima}(2021)}]{shimojima_takahiro_2021_4885407}
\bibinfo{author}{\bibfnamefont{T.}~\bibnamefont{Shimojima}},
  \emph{\bibinfo{title}{Data for science paper 2021}} (\bibinfo{year}{2021}),
  \urlprefix\url{https://doi.org/10.5281/zenodo.4885407}.

\bibitem[{\citenamefont{Klein et~al.}(2020)\citenamefont{Klein, Christensen,
  and Fernandes}}]{PhysRevResearch.2.013336}
\bibinfo{author}{\bibfnamefont{A.}~\bibnamefont{Klein}},
  \bibinfo{author}{\bibfnamefont{M.~H.} \bibnamefont{Christensen}},
  \bibnamefont{and} \bibinfo{author}{\bibfnamefont{R.~M.}
  \bibnamefont{Fernandes}}, \bibinfo{journal}{Phys. Rev. Research}
  \textbf{\bibinfo{volume}{2}}, \bibinfo{pages}{013336} (\bibinfo{year}{2020}).

\bibitem[{\citenamefont{Reiss et~al.}(2017)\citenamefont{Reiss, Watson, Kim,
  Haghighirad, Woodruff, Bruma, Clarke, and Coldea}}]{PhysRevB.96.121103}
\bibinfo{author}{\bibfnamefont{P.}~\bibnamefont{Reiss}},
  \bibinfo{author}{\bibfnamefont{M.~D.} \bibnamefont{Watson}},
  \bibinfo{author}{\bibfnamefont{T.~K.} \bibnamefont{Kim}},
  \bibinfo{author}{\bibfnamefont{A.~A.} \bibnamefont{Haghighirad}},
  \bibinfo{author}{\bibfnamefont{D.~N.} \bibnamefont{Woodruff}},
  \bibinfo{author}{\bibfnamefont{M.}~\bibnamefont{Bruma}},
  \bibinfo{author}{\bibfnamefont{S.~J.} \bibnamefont{Clarke}},
  \bibnamefont{and} \bibinfo{author}{\bibfnamefont{A.~I.}
  \bibnamefont{Coldea}}, \bibinfo{journal}{Phys. Rev. B}
  \textbf{\bibinfo{volume}{96}}, \bibinfo{pages}{121103}
  (\bibinfo{year}{2017}).

\bibitem[{\citenamefont{Chandra and Islam}(2010)}]{CHANDRA20102072}
\bibinfo{author}{\bibfnamefont{S.}~\bibnamefont{Chandra}} \bibnamefont{and}
  \bibinfo{author}{\bibfnamefont{A.}~\bibnamefont{Islam}},
  \bibinfo{journal}{Physica C: Superconductivity}
  \textbf{\bibinfo{volume}{470}}, \bibinfo{pages}{2072} (\bibinfo{year}{2010}).

\bibitem[{\citenamefont{Shilkrot and Srolovitz}(1997)}]{PhysRevB.55.4737}
\bibinfo{author}{\bibfnamefont{L.~E.} \bibnamefont{Shilkrot}} \bibnamefont{and}
  \bibinfo{author}{\bibfnamefont{D.~J.} \bibnamefont{Srolovitz}},
  \bibinfo{journal}{Phys. Rev. B} \textbf{\bibinfo{volume}{55}},
  \bibinfo{pages}{4737} (\bibinfo{year}{1997}).

\bibitem[{\citenamefont{Müller and Saúl}(2004)}]{MULLER2004157}
\bibinfo{author}{\bibfnamefont{P.}~\bibnamefont{Müller}} \bibnamefont{and}
  \bibinfo{author}{\bibfnamefont{A.}~\bibnamefont{Saúl}},
  \bibinfo{journal}{Surface Science Reports} \textbf{\bibinfo{volume}{54}},
  \bibinfo{pages}{157} (\bibinfo{year}{2004}).

\end{thebibliography}

\pagebreak
\widetext
\onecolumngrid

\setcounter{equation}{0}
\setcounter{figure}{0}
\setcounter{table}{0}
\setcounter{page}{1}
\makeatletter

\renewcommand{\theequation}{S\arabic{equation}}
\renewcommand{\thefigure}{S\arabic{figure}}

\begin{large}
\begin{center}
\textbf{Supplementary Material: Defect-induced electronic smectic state at the surface of nematic materials} 
\end{center}  
\end{large}
\FloatBarrier
\section{Defect-generated nematic potential}
In this section, we discuss the properties of the effective nematic potential as a function of $z$ and $z'$. In momentum space (i.e. $q_x$ space), the potential is given by (see Eq. \eqref{reta} of the main text):
\begin{equation}
V_{q_x,z,z'}=\frac{-(g\sigma)^2\beta}{2}q_x^2[|q_x|z+2\nu-1][|q_x|z'+2\nu-1]e^{-|q_x|(z+z')},
\label{retar}
\end{equation}

\begin{figure}
\includegraphics[height=2.45in]{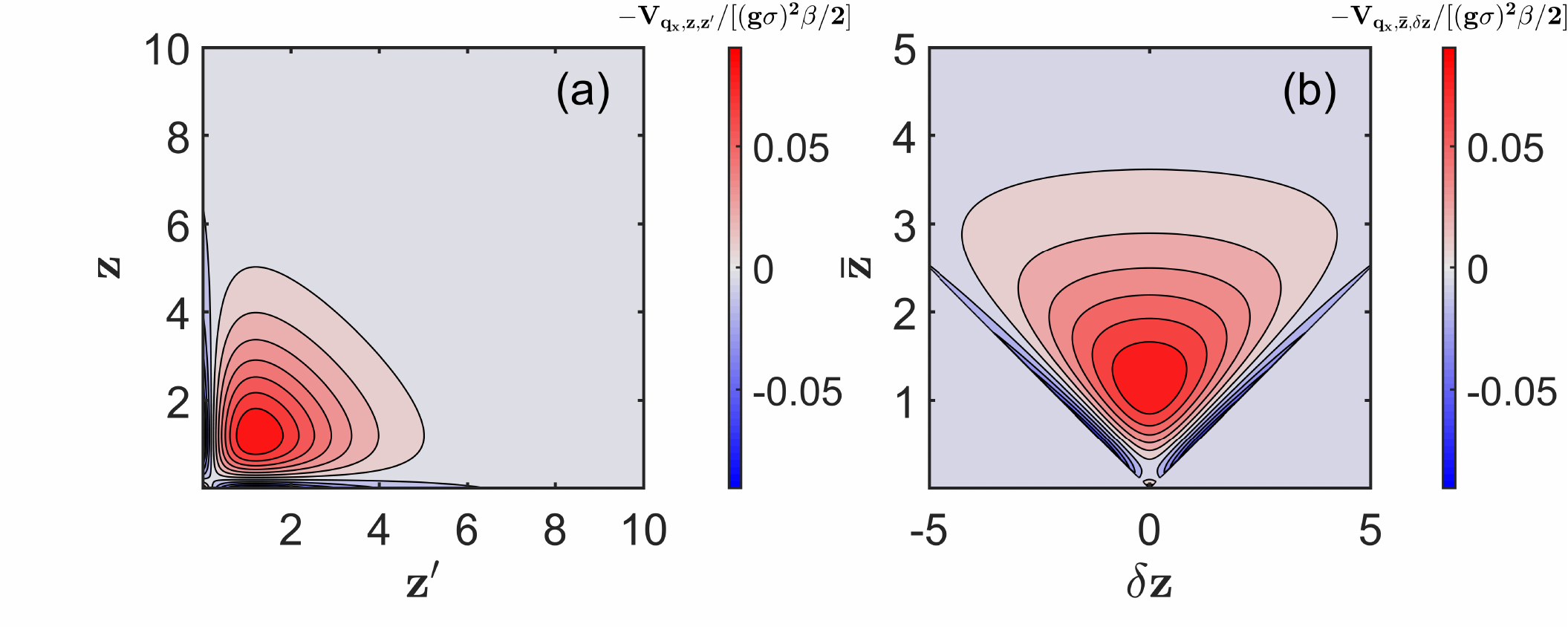}
\caption{The negative of the defect-induced nematic potential, $-V_{q_x,z,z}$, in units of $(g\sigma)^2\beta/2$, and as a function of (a) $z$ and $z'$, and (b) $\bar{z}=(z+z')/2$ and $\delta z = z-z'$. In both panels, $q_x=1$ and $\nu=0.4$. Note that  $-V_{q_x,z,z}$ is peaked at $z=z'\sim 1/q_x$, corresponding to $\bar{z}\sim 1/q_x$ and $\delta z=0$.}
\label{crfig1}
\end{figure}

For a fixed $q_x$, this function, shown in Fig. \ref{crfig1} for $\nu=0.4$ and $q_x = 1$, is characterized by a well-defined negative minimum centered around $z, z' \sim 1/q_x$ (corresponding to a positive peak of $-V_{q_x,z,z'}$). The width of this minimum is approximately the same along the directions $z$, $z'$ and $z-z'$. As a result, when the defect-induced potential is rewritten in terms of the quantities $\bar{z}=(z+z')/2$ and $\delta z=z-z'$, as shown in Fig. \ref{crfig1}(b), the minimum has approximately the same width along both $\bar{z}$ and $\delta z$ coordinate directions. In terms of these coordinates, the minimum is centered at $\delta z = 0$ and $\bar{z} \sim 1/q_x$.

It is also convenient to study the potential in the real $x$-space. Performing a Fourier transform of Eq. \eqref{retar}, we obtain:
\begin{align}
V_{\delta x,z,z'}&=\frac{-(g\sigma)^2\beta}{2}\frac{L_\parallel}{(2\pi)}\bigg[\frac{48 zz'}{(z+z')^5}\frac{1-10\xi^2+5\xi^4}{(\xi^2+1)^5} + \frac{12(2\nu-1)}{(z+z')^3}\frac{1-6\xi^2+\xi^4}{(\xi^2+1)^4}+ \frac{4(2\nu-1)^2}{(z+z')^3}\frac{1-3\xi^2}{(\xi^2+1)^3}\bigg]\nonumber\\
&\equiv \frac{-(g\sigma)^2\beta}{2}\frac{L_\parallel}{(2\pi)}\psi\bigg(z,z',\xi=\frac{\delta x}{z+z'},\nu\bigg).\label{Cstepr1}
\end{align}

\begin{figure}
\includegraphics[height=2.28in]{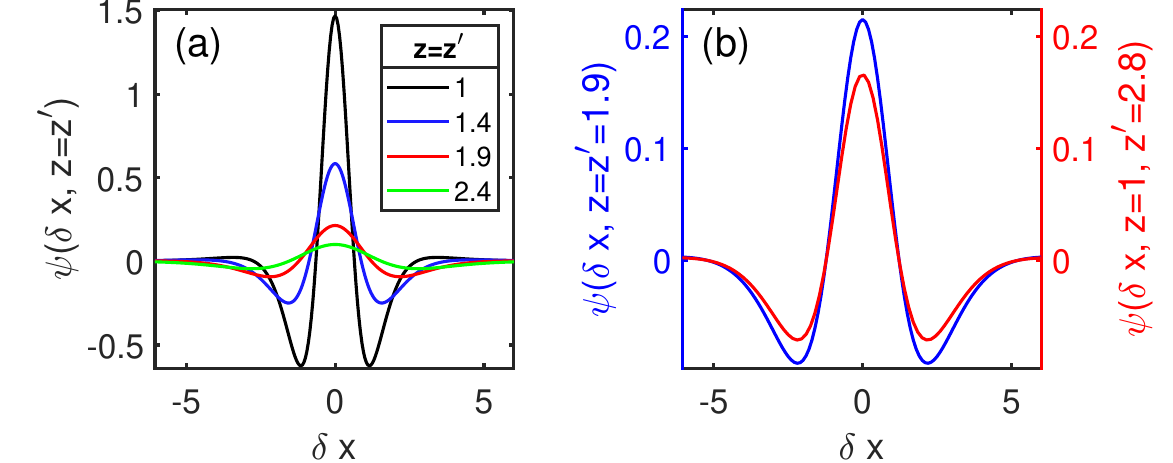}
\caption{The auxiliary real-space function $\psi \propto -2V/(g\sigma)^2\beta$ in Eq. \eqref{Cstepr1}, plotted as a function of $\delta x$, for the cases of (a) increasing mean depth $(z+z')/2$ but $z=z'$, and (b) increasing depth difference $z-z'$ but same mean depth $(z+z')/2$. Here, we set $\nu=0.4$.} 
\label{crfig}
\end{figure}

Consistent with the analysis in momentum $q_x$-space, the relevant range of $\delta x$ is of the order of $(z+z')$, corresponding to the momentum scale $q_x\sim 1/(z+z')$. This is illustrated by the behavior of the auxiliary function $\psi$ shown in Fig. \ref{crfig}. In panel (a), we note that the the peak-to-trough distance increases with increasing mean depth $(z+z')/2$. More importantly, the change in sign of $\psi$ over this distance makes the values of the nematic order parameter at points separated by this distance to also have opposite signs, thereby generating a modulation. Panel (b) demonstrates that two $\psi$ curves with the same mean depth $(z+z')/2$ but different $z-z'$ have a very similar shape.

\section{Smectic critical temperature for step defects}
\label{smectc}
Here we derive the smectic critical temperature $T_{\mathrm{smc}}$ by minimizing the linearized disorder-averaged action. Using Eq. \eqref{eq:S-d} of the main text, the disorder-averaged nematic action is given by,
\begin{align}
S=&L_\parallel^2\sum_{q_x}\int_{z=0}^L \bigg[\bigg(r_0\frac{T-T_{\text{nem}}}{2T_{\text{nem}}^{(0)}}\bigg)|\eta_{q_x,z}|^2+\frac{b_\parallel}{2}q_x^2|\eta_{q_x,z}|^2+\frac{b}{2}|\partial_z\eta_{q_x,z}|^2-\int_{z'=0}^L\frac{(g\sigma)^2\beta}{2}q_x^2[|q_x|z+2\nu-1][|q_x|z'+2\nu-1]\nonumber\\ 
&\hspace*{15.5mm}\times e^{-|q_x|(z+z')}\eta_{-q_x,z'}\eta_{q_x,z}\bigg],
\end{align}
Defining the constant parameter $R=\int_{z'}|q_x|[|q_x|z'+2\nu-1]e^{-|q_x|z'}\eta_{-q_x,z'}$, the linearized saddle-point equation is given by, 
\begin{equation}
\begin{split}
&\partial_z^2\eta_{q_x,z}-\frac{\big(r_0\frac{T-T_{\text{nem}}}{T_{\text{nem}}^{(0)}}+b_\parallel q_x^2\big)}{b}\eta_{q_x,z}+(g\sigma)^2 \beta \frac{R |q_x| (|q_x|z+2\nu-1)e^{-|q_x|z}}{b}=0.
\end{split}
\end{equation}
For a sample occupying the half-space $z\geq 0$, its solution is readily obtained as,
\begin{align}
\eta_{q_x,z}^{(\text{sp})}=&\frac{(g\sigma)^2 \beta }{b }\frac{e^{-\sqrt{\frac{(t+b_\parallel q_x^2)}{b}}z} \Big[\frac{2 (t+b_\parallel q_x^2) q_x^2 (1-\nu)}{b}+2\nu q_x^4 \Big]}{\sqrt{\frac{(t+b_\parallel q_x^2)}{b}} \Big[\frac{(t+b_\parallel q_x^2)}{b}-q_x^2\Big]^2}R+\frac{(g\sigma)^2 \beta }{b}\frac{e^{-|q_x| z} \Big[\frac{(t+b_\parallel q_x^2-bq_x^2)}{b} |q_x| (|q_x|z+2\nu-1)-2 |q_x|^3\Big]}{ \Big[\frac{(t+b_\parallel q_x^2)}{b}-q_x^2\Big]^2}R,\label{ansolbneq0}
\end{align}

\begin{figure}
\centering
\includegraphics[height=2.07in]{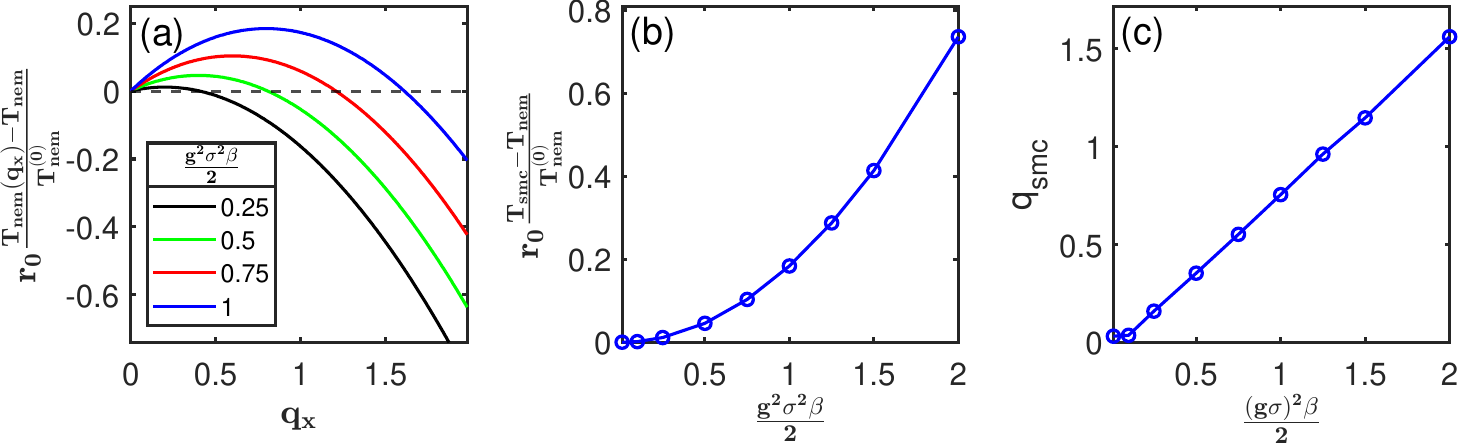}
\caption{(a) The smectic critical temperature as a function of $q_x$, obtained using Eq. \eqref{ansolbneq0}. (b) The expected smectic critical temperature, obtained from the peak values in the left panel, is found to vary quadratically with the effective defect strength $(g\sigma)^2\beta /2$. (c) The smectic wavevector $q_{\text{smc}}$ as a function of the effective defect strength $(g\sigma)^2\beta /2$, is found to increase approximately linearly.}
\label{tsmecq}
\end{figure}

In this expression, we defined $t=r_0(T-T_{\text{nem}})/T_{\text{nem}}^{(0)}$ for brevity and used the von Neumann boundary condition $\partial_z \eta (z\rightarrow0) = 0$ since no nematic surface terms are present. The smectic critical temperature for a given wave-vector $q_x$ is obtained from the self-consistency condition, $R=\int_{z'=0}^L |q_x|(|q_x|z'+2\nu-1)e^{-|q_x|z'}\eta^{(\text{sp})}_{-q_x, z'}$, as shown in Fig. \ref{tsmecq}(a). As expected, the reduced critical temperature vanishes for $q_x=0$. Upon increasing $q_x$, it rises due to the defect contribution to the potential, and is eventually peaked at a finite $q_x$, followed by a suppression caused by the nematic stiffness contribution to the potential. The actual smectic critical temperature, obtained from the peak values in Fig. \ref{tsmecq}(a), is shown in Fig. \ref{tsmecq}(b). It is found to increase quadratically with the effective defect strength $(g\sigma)^2\beta /2$. The observed smectic wave-vector, corresponding to the peak positions in Fig. \ref{tsmecq}(a), is shown in Fig. \ref{tsmecq}, and varies approximately linearly with the defect strength. (c). 

An analytical approximation may be obtained in the limit $b\to 0$. This is a reasonable approximation for layered materials, such as the iron-based superconductors. In this limit, the saddle-point solution is given by,
\begin{align}
\eta^{(\text{sp})}_{q_x,z}=(g\sigma)^2 \beta \frac{ |q_x| [|q_x|z+2\nu-1]e^{-|q_x|z}R}{\big(r_0\frac{T-T_{\text{nem}}}{T_{\text{nem}}^{(0)}}+b_\parallel q_x^2\big)}, \label{aux_1}
\end{align}
from which the critical temperature is obtained by imposing the same self-consistency condition as before, 
\begin{align}
r_0\frac{T_{\text{nem}}(q_x)-T_{\text{nem}}}{T_{\text{nem}}^{(0)}}&=(g\sigma)^2 \beta  \bigg[\bigg(\nu-\frac{1}{2}\bigg)^2+\nu^2\bigg]|q_x| -b_\parallel q_x^2.\label{acqxanb0}
\end{align}

It is clear that the maximum transition temperature happens at a non-zero $q_x$, leading to the wave-vector:

\begin{equation}
|q_{\text{smc}}|=\frac{(g\sigma)^2\beta }{2b_\parallel}\big[(\nu-\frac{1}{2})^2+\nu^2\big]
\end{equation}

The smectic critical temperature is obtained by substituting $T_{\text{smc}}=\max_{q_x}T_{\text{nem}}(q_x)$,
\begin{align}
r_0\frac{T_{\text{smc}}-T_{\text{nem}}}{T_{\text{nem}}^{(0)}}&=\frac{(g\sigma)^4 \beta^2 }{4b_\parallel}\bigg[\bigg(\nu-\frac{1}{2}\bigg)^2+\nu^2\bigg]^2=b_\parallel q_{\text{smc}}^2. \label{eq:aux1}
\end{align}

Furthermore, from the spatial profile of the nematic order parameter, Eq. \ref{aux_1}, we note that $\eta_{q_x,z}$ is exponentially localized on the surface and peaked at $z\sim 1/|q_{\text{smc}}|$.

\section{Smectic order in the case of point defects}\label{pointdef}
\begin{figure}[htb!]
\centering
\includegraphics[height=1.7in]{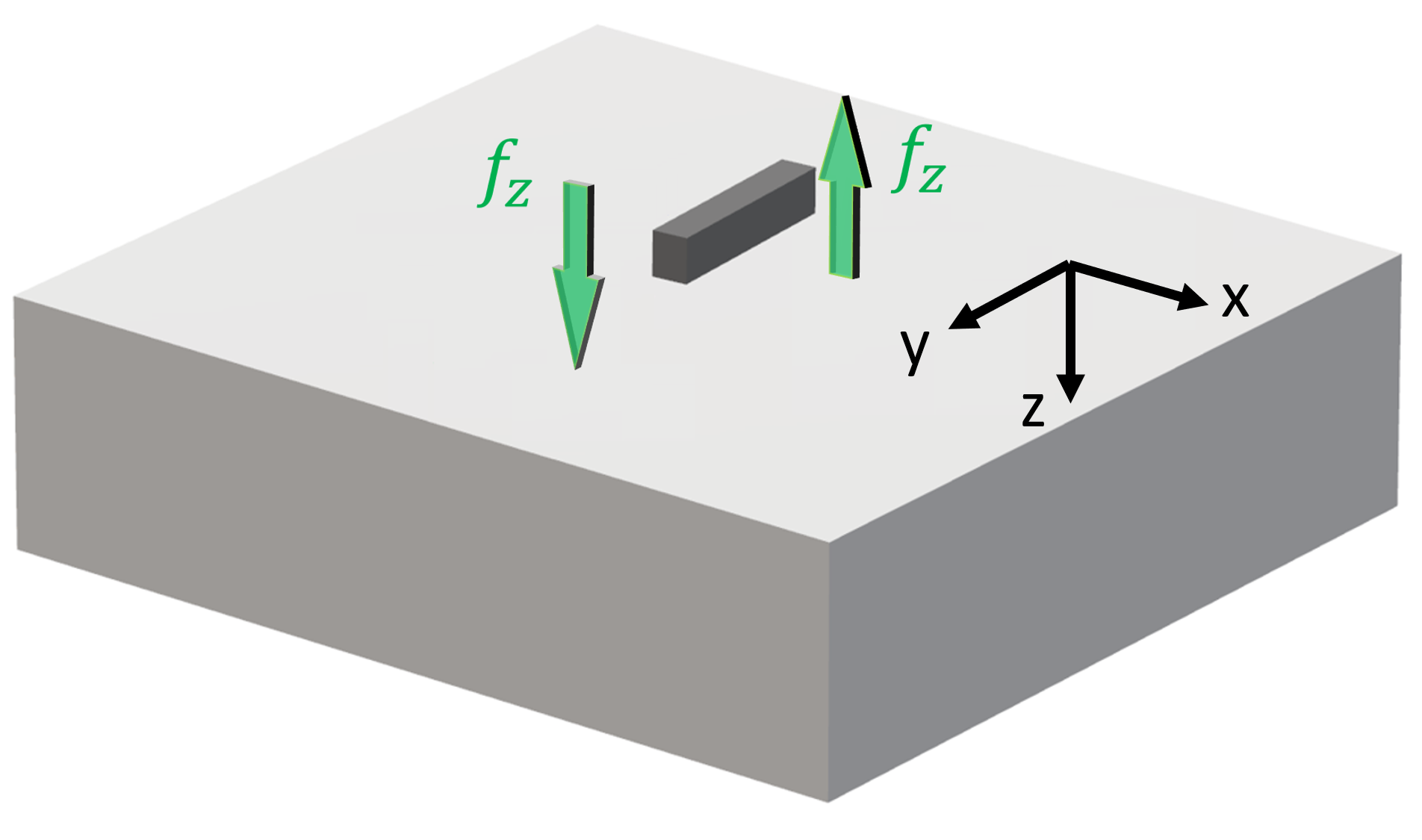}
\caption{A generic illustration of the forces generated by a point-like anisotropic defect, parametrized by $f^{(1)}_z=h^{(1)}_z\partial_x\delta(x-x')\delta(y-y')$.}
\label{Anptfig}
\end{figure}

In this section, we study the case where electronic smecticity is induced not by step-like defects, but by point-like anisotropic defects illustrated in Fig. \ref{Anptfig}. We follow the same procedure as in the case of infinite steps and start by obtaining the strain field for a point-like anisotropic defect. We subsequently obtain the defect-induced nematic potential by averaging over a distribution of such defects.

The strain created by a defect in equilibrium is modeled by a localized force density $f_\mu\sim \partial_x^m\delta(x-x')\partial_y^n\delta(y-y')$ where $\mu=x,y$ denotes the $x,y-$directions, and $(x',y')$ denotes the location of the defect \cite{MarchenkoParshin,PhysRevB.53.11120,PhysRevB.49.13848,doi:https://doi.org/10.1002/352760667X.ch9,BACON198051,CLOUET201849} . Since defects in equilibrium cannot produce a net force, $m+n>0$. We restrict ourselves to defects described by a dipolar force along only one direction, as it is the leading order contribution to the strain over long distances. Considering defects aligned with the crystallographic axes, this leads to two possibilities for the defect force densities, namely, $f^{(1)}_\mu=h^{(1)}_\mu\zeta[\partial_x\delta(x-x')]\delta(y-y')\delta(z)$ and $f^{(2)}_\mu=h^{(2)}_\mu\zeta\delta(x-x')[\partial_y\delta(y-y')]\delta(z)$, along with superpositions of these two forces. Here $h^{(1,2)}$ denote the corresponding forces, and $\zeta\approx a_\parallel$ is a microscopic length scale of the order of the lattice constant. The first case, $(f_\mu^{(1)})$, is depicted in Fig. \ref{Anptfig}. 

Now, we derive the $B_{1g}$ strain field generated by the defect force $f_\mu^{(1)}$. The strain field corresponding to $f_\mu^{(2)}$ is obtained by interchanging $x\leftrightarrow y$ and $y\leftrightarrow-x$. As in the main text, for simplicity, we restrict ourselves to the case where only the $z-$component of the force is present, i.e. $f_{x,y}^{(1)}=0$, but $f_z^{(1)}\equiv f^{(1)}\neq 0$. The Green's function for a unit point force density $\delta(x)\delta(y)\delta(z)$ applied normally to the surface of a semi-infinite elastic half-space $(z\geq 0)$ at the origin is given by \cite{LIFSHITZ198687}
\begin{align}
G_{jz}( x,y,z)=&\frac{(1+\nu)}{2\pi E}\bigg[\frac{zx_{j}}{\scriptr^{3}}+(3-4 \nu)\frac{\delta_{j,3}}{\scriptr}-\frac{(1-2\nu)}{\scriptr+z}\bigg(\delta_{j,3}+\frac{x_j}{\scriptr}\bigg)\bigg],
\end{align}
where $\scriptr=\sqrt{x^2+y^2+z^2}$. Hence, the deformation profile created by this single defect at the origin is given by,
\begin{align}
u_x^{(1)}=&h^{(1)}\zeta\partial_xG_{xz}(x,z)=\frac{(1+\nu)\zeta h^{(1)}}{2\pi E}\partial_x\bigg[\frac{zx}{\scriptr^{3}}-\frac{(1-2\nu)}{\scriptr+z}\frac{x}{\scriptr}\bigg]\approx\frac{(1+\nu)\zeta h^{(1)}}{2\pi E}\partial_x\bigg[\frac{zx}{\scriptr^{3}}-(1-2\nu)\frac{x}{\scriptr^2}\bigg],\\
u_y^{(1)}=&h^{(1)}\zeta\partial_xG_{yz}(x,z)=\frac{(1+\nu)\zeta h^{(1)}}{2\pi E}\partial_x\bigg[\frac{zy}{\scriptr^{3}}-\frac{(1-2\nu)}{\scriptr+z}\frac{y}{\scriptr}\bigg]\approx\frac{(1+\nu)\zeta h^{(1)}}{2\pi E}\partial_x\bigg[\frac{zy}{\scriptr^{3}}-(1-2\nu)\frac{y}{\scriptr^2}\bigg]
\end{align}
Here, we have approximated the denominator $\scriptr+z\approx \scriptr$ to obtain an analytical closed-form expression in Fourier domain. This is is valid over long distances $\scriptr$, corresponding to $q_\parallel\to 0$. Defining $r=\sqrt{x^2+y^2}$, $x=r\cos\theta$, $y=r\sin\theta$, and $\bm{q}_\parallel=(q_x,q_y)=q_\parallel(\cos\phi,\sin\phi)$, along with $\omega=q_\parallel r$ and $\gamma=q_\parallel z$, we have the following Fourier transformed deformations,
\begin{align}
u_{x,\bm{q}_\parallel}^{(1)}\approx&\frac{(1+\nu)\zeta h^{(1)}}{2\pi E}\frac{1}{L_\parallel^2}\big[\gamma e^{-\gamma}-(1-2\nu)\gamma K_{1,\gamma}\big]\cos^2(\phi),\\
u_{y,\bm{q}_\parallel}^{(1)}\approx&\frac{(1+\nu)\zeta h^{(1)}}{2\pi E}\frac{1}{L_\parallel^2}\big[\gamma e^{-\gamma}-(1-2\nu)\gamma K_{1,\gamma}\big]\cos(\phi)\sin(\phi),
\end{align}
where $K_{n,z}$ is the modified Bessel function of the second kind. Therefore, the $B_{1g}$ strain for a defect with unit force $h^{(1)}=1$, $\bar{\varepsilon}_{B_{1g}}=(\partial_xu_x-\partial_yu_y)/\sqrt{2}\vert_{h^{(1)}=1}$, is given by
\begin{align}
\bar{\varepsilon}_{1,\bm{q}_\parallel,z}^{B_{1g}}\approx &\frac{(1+\nu)\zeta }{2\sqrt{2}\pi E}\frac{1}{L_\parallel^2}(-iq_\parallel)\big[\gamma e^{-\gamma}-(1-2\nu)\gamma K_{1,\gamma}\big]\big[\cos^3(\phi)-\cos(\phi)\sin^2(\phi)\big].\label{strainm1}
\end{align}

To obtain the defect-induced nematic potential, we follow the procedure presented in the main text for the case of step defects. Each defect, randomly distributed and indexed by $j$, is located at the sample surface ($z=0$) with the location specified by $\bm{r}_{\parallel,j}=(x_j,y_j)$ and aligned along one of the crystallographic axes, $(m)=\{(1),(2)\}$. The net strain created by the random distribution of defects can be written as $\varepsilon^{B_{1g}}_{\bm{r}}=\sum_j\big[h^{(1)}_j\bar{\varepsilon}^{B_{1g}}_{1,\bm{r}-\bm{r}_j}+h^{(2)}_j\bar{\varepsilon}^{B_{1g}}_{2,\bm{r}-\bm{r}_j}\big]$. It is convenient to define the defect force density
\begin{align}
\rho^{(m)}_{\bm{r}}=\sum_j h^{(m)}_j\delta(x-x_{j})\delta(y-y_{j}),
\end{align} 

Similar to the step defects considered in the main text, the variance of this continuous defect distribution is $\big(\sigma_{m}\big)^2\frac{N_{\text{def}}}{L_\parallel^2}\frac{a_\parallel^2}{L_\xi^2}$, with $N_{\text{def}}$ being the total number of defects, and $L_\xi$, a length scale larger than $a_\parallel$ but smaller than the bare nematic correlation length (as introduced in the main text). We thus obtain the action, 
\begin{align}
S=&\int_{\bm{r}_\parallel,z}\eta_{\bm{r}_\parallel,z}\frac{r_0\frac{T-T_{\text{nem}}}{T_{\text{nem}}^{(0)}}-b_\parallel\nabla_\parallel^2-b\partial_z^2}{2}\eta_{\bm{r}_\parallel,z}-g\sum_{m=1,2}\int_{\bm{r}_\parallel,z}\int_{\bm{r}''_\parallel}\rho^{(m)}_{\bm{r}''_\parallel}\bar{\varepsilon}^{B_{1g}}_{m,\bm{r}-\bm{r}''_\parallel}\eta_{\bm{r}_\parallel,z}.
\end{align}
After averaging over the defect distributions, we find
\begin{align}
S=&\int_{\bm{r}_\parallel,z}\eta_{\bm{r}_\parallel,z}\frac{r_0\frac{T-T_{\text{nem}}}{T_{\text{nem}}^{(0)}}-b_\parallel\nabla_\parallel^2-b\partial_z^2}{2}\eta_{\bm{r}_\parallel,z}+\sum_{m=1,2}\frac{\big(g\sigma_{m}\big)^2n_{\text{def}}\frac{L_\xi^2}{a_\parallel^2}}{2}\int_{\bm{r}''_\parallel}\int_{\substack{\bm{r}_\parallel',z'\\ \bm{r}_\parallel,z}}\bar{\varepsilon}^{B_{1g}}_{m,x-x'',y,z}\bar{\varepsilon}^{B_{1g}}_{m,x'-x'',y',z'}\eta_{\bm{r}_\parallel,z}\eta_{\bm{r}_\parallel',z'}\nonumber\\
=&L_\parallel^2\int_{z}\sum_{\bm{q}_\parallel}\eta_{-\bm{q}_\parallel,z}\frac{r_0\frac{T-T_{\text{nem}}}{T_{\text{nem}}^{(0)}}+b_\parallel q_\parallel^2-b\partial_z^2}{2}\eta_{\bm{q}_\parallel,z}-L_\parallel^2\int_{z,z'}\sum_{\bm{q}_\parallel}\underbrace{\sum_{m=1,2}\frac{-\big(g\sigma_{m}\big)^2\beta}{2}F_{m,\bm{q}_\parallel,z}F_{m,-\bm{q}_\parallel,z'}}_{V_{\bm{q}_\parallel,z,z'}}\eta_{-\bm{q}_\parallel,z}\eta_{\bm{q}_\parallel',z'}\Bigg],\label{zrt}
\end{align}

where, from Eq. \eqref{strainm1}, $F_{m,\bm{q}_\parallel,z}=L_\parallel^2\bar{\varepsilon}^{B_{1g}}_{m,\bm{q}_\parallel,z}/\big(\frac{(1+\nu)\zeta }{2\sqrt{2}\pi E}\big)\approx (-iq_\parallel)\big[\gamma e^{-\gamma}-(1-2\nu)\gamma K_{1,\gamma}\big]\big[\cos^3(\phi_{m})-\cos(\phi_{m})\sin^2(\phi_{m})\big]$ with $=\phi+(m-1)\frac{\pi}{2}$. The last expression defines the analogue of $V_{q_x,z,z'}$ defined by Eq. \eqref{reta} in the main text, but for the point defects being considered here. The effective coupling in Fourier space is then given by $\big(g\sigma_{m}\big)^2\beta/2$, with $\beta=\big[N_{\text{def}} (L_\xi^2/a_\parallel^2)\big] \big[(1+\nu)\zeta/(2\sqrt{2}\pi E)\big]^2$.
As a result, the defect-induced potential is explicitly given by
\begin{align}
V_{\bm{q}_\parallel,z,z'}\approx&\sum_{m=1,2}\frac{-(g\sigma_{m})^2\beta }{2}q_\parallel^2\big[q_\parallel z e^{-q_\parallel z}-(1-2\nu)q_\parallel z K_{1,q_\parallel z}\big]^2\big[\cos^3(\phi_{m})-\cos(\phi_{m})\sin^2(\phi_{m})\big]^2.\label{retafd}
\end{align}

\begin{figure}[htb!]
\centering
\includegraphics[height=3in]{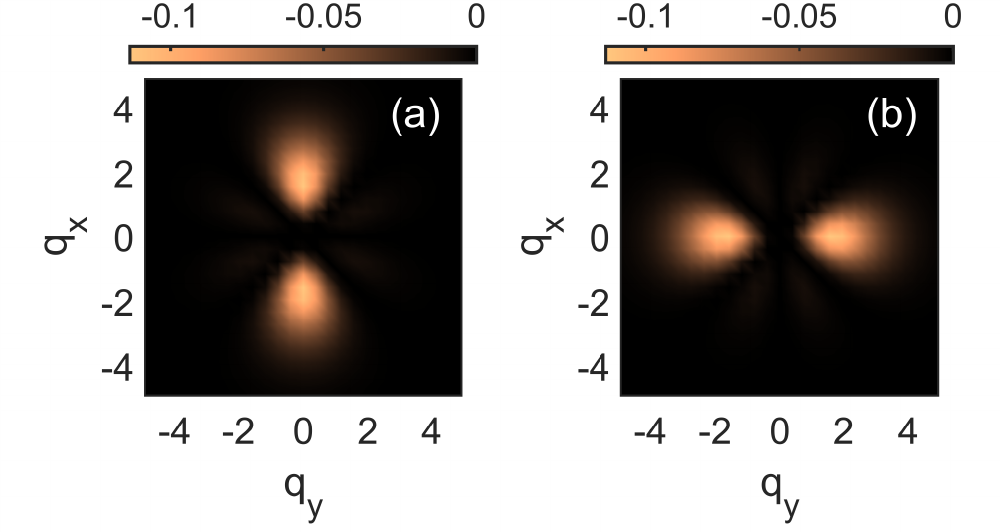}
\caption{The analytical approximation to the defect-generated potential $V_{\bm{q}_\parallel,z,z'}/\frac{(g\sigma_{m})^2\beta }{2}$, Eq. \eqref{retafd}, for (a) $m=1$ and (b) $m=2$, considering $z=z'$. In (a), troughs appear at $q_x\neq 0$, $q_y=0$ while in (b), troughs are found at $q_y\neq 0$, $q_x=0$.}
\label{anisoptdef}
\end{figure} 

This potential, plotted in Fig. \ref{anisoptdef}, has a trough at finite $q_\parallel$ and, hence, it favors a modulated nematic order. Note that, depending on the character of the defect distribution, $\sigma_{1}$ and $\sigma_{2}$ could be different. In this case, the nematic potential would have deeper troughs along one direction than the other, leading to stripe smectic patterns.

To estimate the smectic wave-vector, we write the nematic potential in Eq. \eqref{retafd} terms of the variables $\bar{z}=(z+z')/2$ and $\delta z=z-z'$. Similarly to the case of step defects, the (negative) potential is peaked at $\bar{z}\sim 1/q_\parallel$ and $\delta z=0$, leading to a nematic order parameter $\eta_{q_\parallel,z}$ that is peaked near the surface $(z=0)$. Assuming that $\eta_{q_\parallel,z}$ varies slowly near the surface over a depth $L_s\sim 1/q_\parallel$, and eventually decays exponentially away from the surface, the integral in Eq. \eqref{eq:S-d} yields,
\begin{equation}
\begin{split}
S_d = &L_\parallel^2\int_{\bar{z}=-L}^0\int_{\delta z=-L_s/2}^{L_s/2}\sum_{\bm{q}_\parallel} V_{\bm{q}_\parallel,\bar{z},\delta z} |\eta_{\bm{q}_\parallel,0}|^2\nonumber\\
\approx&L_\parallel^2 L_s\sum_{\bm{q}_\parallel} \sum_{m=1,2}\frac{-(g\sigma_{m})^2\beta }{2}q_\parallel |\eta_{\bm{q}_\parallel,0}|^2 \int_{\omega=0}^{\infty} d\omega \big[\omega e^{-\omega}-(1-2\nu)\omega K_{1,\omega}\big]^2\big[\cos^3(\phi_{m})-\cos(\phi_{m})\sin^2(\phi_{m})\big]^2,\nonumber\\
=&L_\parallel^2 L_s\sum_{\bm{q}_\parallel} \sum_{m=1,2}\frac{-(g\sigma_{m})^2\beta }{2}q_\parallel d(\nu,\phi,m)|\eta_{q_x,0}|^2,
\end{split}
\end{equation}
where,
\begin{align}
d(\nu,\phi,m)=&\bigg[\frac{1}{4}+(1-2\nu)^2\frac{3\pi^2}{32}-4\frac{(1-2\nu)}{5}\bigg]\big[\cos^3(\phi_{m})-\cos(\phi_{m})\sin^2(\phi_{m})\big]^2.
\end{align}

Recall that $\phi_{m}=\phi+(m-1)\pi/2$. In the limit $b\to 0$, the minimization of $\mathcal{S}\approx L_\parallel^2 L_s[(T-T_{\text{nem}})/2T_{\text{nem}}+b_\parallel q_{\text{smec}}^2/2]|\eta_{\bm{q}_{\text{smec}},0}|^2+\mathcal{S}_d$ takes place for a non-zero $q_{\text{smec}} = \sum_{m=1,2}(g\sigma_{m})^2\beta d(\nu,\phi)/2b_\parallel$, with $\phi=n\pi/2$ and $n\in \mathbb{Z}$, and at the temperature $T_{\text{smec}}= T_{\text{nem}}(1 + b_\parallel q_{\text{smec}}^2)$.

\section{Smectic-Nematic transition}
Here we study the smectic-nematic phase transition, focusing on the nature of the transition as well as on its dependence on the sample thickness. There are two competing factors affecting the transition. On the one hand, the defect-induced potential makes the smectic mass (i.e. the coefficient of the quadratic term in the action) more negative than  the uniform nematic mass, which favors smectic order. On the other hand, the extensive character of the bulk-nematic free energy gain favors the bulk nematic order at low temperatures.

The quartic term $(1/4)u_\eta\eta_{\bm{r}}^4$, when written in momentum space, is equivalent to a biquadratic ``repulsion" $\sum_{\bm{q},\bm{q}'}(1/4)u_\eta |\eta_{\bm{q}}|^2|\eta_{\bm{q}'}|^2$. Thus, for any temperature, it selects the order with the lowest mass. Consequently, for the transition to the bulk-nematic phase to occur, the extensive nematic free energy gain must compensate for the defect-induced enhancement of the non-extensive surface-smectic free energy. From Eq. \eqref{S_nem} in the main text, the free energy can be written as
\begin{align}
F=&F_{\text{nem}}^{(2)}+F_{\text{smc}}^{(2)}+F_{\text{smc-nem}}^{(4)} \, ,
\end{align}
where we neglect all $q$ wave-vectors except the homogeneous nematic wave-vector $q=0$ and the preferred smectic wave-vector $q_{\mathrm{smc}}$. The quadratic part of the nematic free energy, 
\begin{align}
F_{\text{nem}}^{(2)}=&L_\parallel^2L r_0\,\frac{T-T_{\text{nem}}}{2T_{\text{nem}}^{(0)}}\eta_{q_x=0}^2,
\end{align} 
scales extensively with the system size. The surface smectic free energy,
\begin{align}
  F_{\text{smc}}^{(2)}
=&L_\parallel^2 L_s\bigg[\bigg(r_0\frac{T-T_{\mathrm{nem}}}{2T_{\mathrm{nem}}^{(0)}}\bigg)-\frac{b_{\parallel}q_{\text{smc}}^{2}}{2}\bigg]|\eta_{\bm{q}_{\text{smc}},0}|^{2},
\end{align} 
does not scale extensively along the $z-$direction, as the smectic layer is restricted to a depth $L_s\sim1/q_{\text{smc}}$,
where
$q_{\mathrm{smc}}=(g\sigma)^{2}\beta\big[(\nu-\frac{1}{2})^{2}+\nu^{2}\big]/2b_{\parallel}$ as derived in the main text. Lastly, the quartic contribution to the free energy is given by
\begin{align}
F_{\text{smc-nem}}^{(4)}=&L_\parallel^2L_s\bigg[\frac{12u_\eta}{4}|\eta_{q_{\text{smc}}}|^2\eta_{q_x=0}^2+\frac{6u_\eta}{4}|\eta_{q_{\text{smc}}}|^4\bigg]+L_\parallel^2L\frac{u_\eta}{4}\eta_{\bm{q}_\parallel=0}^4.
\end{align} 

The defect-induced strain $\varepsilon^{B_{1g}}_{\bm{q}_\parallel}$ decays over a depth $z\sim 1/q_\parallel$. In a sufficiently thick sample with $L\gg 1/q_{\text{smc}}$, the description of the system in terms of an elastic half-space is valid. Then, the smectic-nematic transition temperature $T_{\text{smc-nem}}$ can be obtained by equating the masses of the bulk nematic and of the surface smectic free energies, yielding,
\begin{align}
r_0\frac{T_{\mathrm{smc-nem}}-T_{\mathrm{nem}}}{T_{\mathrm{nem}}^{(0)}}=&\frac{-b_{\parallel}q_{\text{smc}}^{2}}{\frac{L}{L_s}-1}<0.\label{smc_nem_bdy1}
\end{align}
Clearly, while the phase boundary approaches $T_{\text{smc-nem}}=T_{\text{nem}}$ for $L\to\infty$, as depicted in Fig.\ref{fig_phase_diagram}, it moves below and farther away from $T_{\text{nem}}$ with decreasing $L$. Since the smectic wave-vector $q_{\text{smc}}$ is expected to remain largely unchanged with varying $L/L_\parallel$, the suppression of $T_{\text{smc-nem}}$ below $T_{\text{nem}}$ should persist for even thinner samples. 

Note that, when the transition occurs at $T_{\text{smc-nem}}=T_{\text{nem}}$, which happens for $L\rightarrow \infty$, the nematic order parameter develops continuously from zero, whereas the smectic field discontinuously drops to zero, $\eta_{q_{\text{smc}}}(T=T_{\text{nem}}^-)=0$. However, when $L$ becomes comparable to $1/q_{\text{smc}}$, such that $T_{\text{smc-nem}}<T_{\text{nem}}$, the nematic field discontinuously jumps to $\eta_{\bm{q}_\parallel=0}(T=T_{\text{smc-nem}}^-)=[(T_{\text{nem}}-T_{\text{smc-nem}})/u_\eta T_{\text{nem}}]^{1/2}$, in conjunction with the smectic field discontinuously dropping to zero.

\section{Comparison with the experimental PEEM data}
In this section, we compare our results with the PEEM data published in Ref. \citen{Shimojima2021}. We first address the puzzling
result
that the smectic state survives deep into the temperature range where a homogeneous nematic phase is expected. We also use the experimental results to estimate the typical energy scale associated with the defects, which we find to be reasonable. This energy scale could in principle be compared to microscopic calculations as a further verification, but that is beyond the scope of the current work.

\begin{figure}[htb!]
\centering
\includegraphics[height=4in]{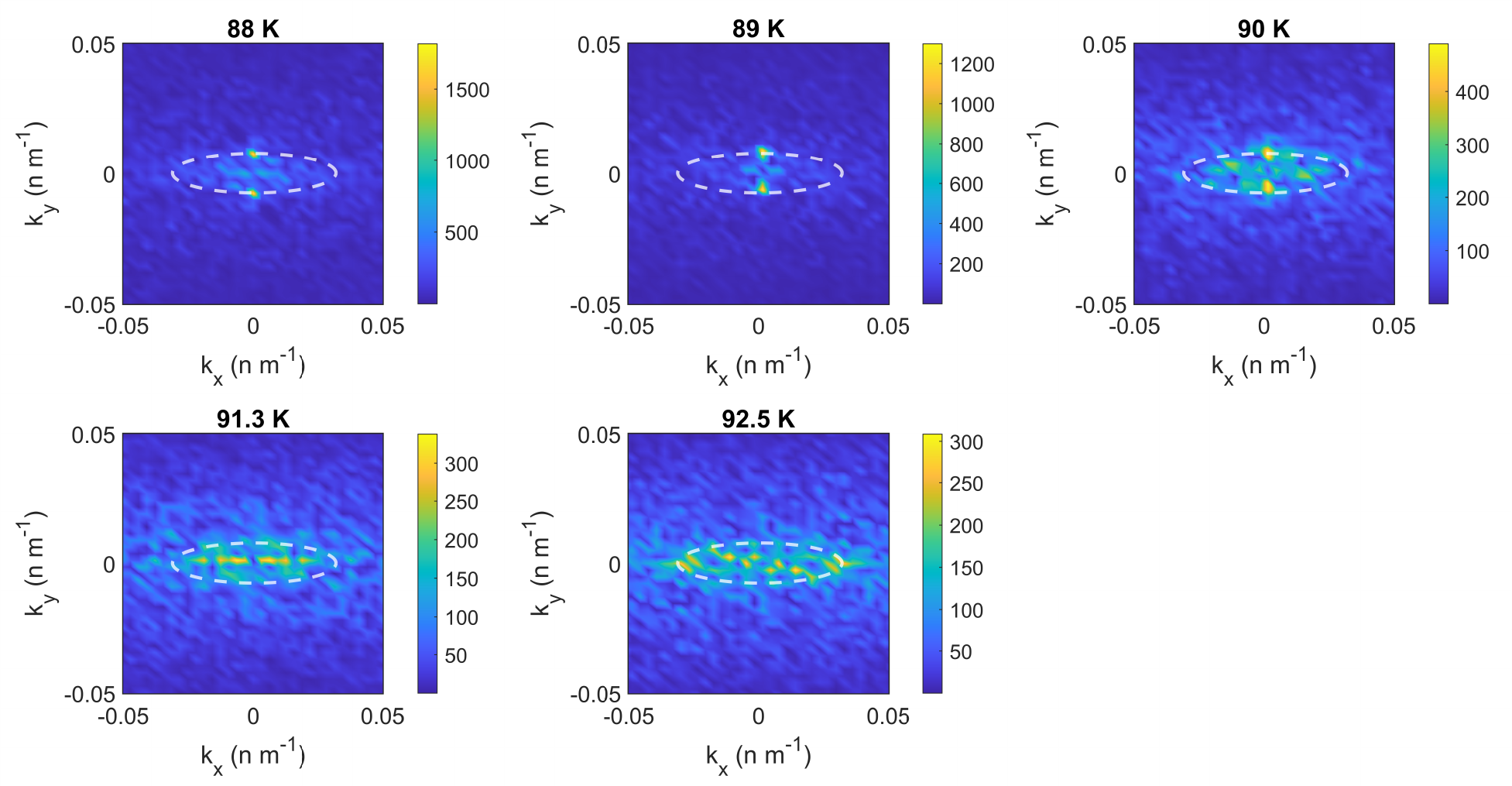}
\caption{The LD PEEM data for FeSe in momentum space. The dashed white line, which is a guide to the eyes, marks the ``spot" of fixed size $\sim q_{\text{smc}}$. Such a spot is more clearly seen in Fig. \ref{FeSeFourierclip}. This plot was generated from the real-space experimental data of Ref. \citen{Shimojima2021}, which is available at Zenodo  \cite{shimojima_takahiro_2021_4885407}.}
\label{FeSeFourier}
\end{figure}

\begin{figure}[htb!]
\centering
\includegraphics[height=4in]{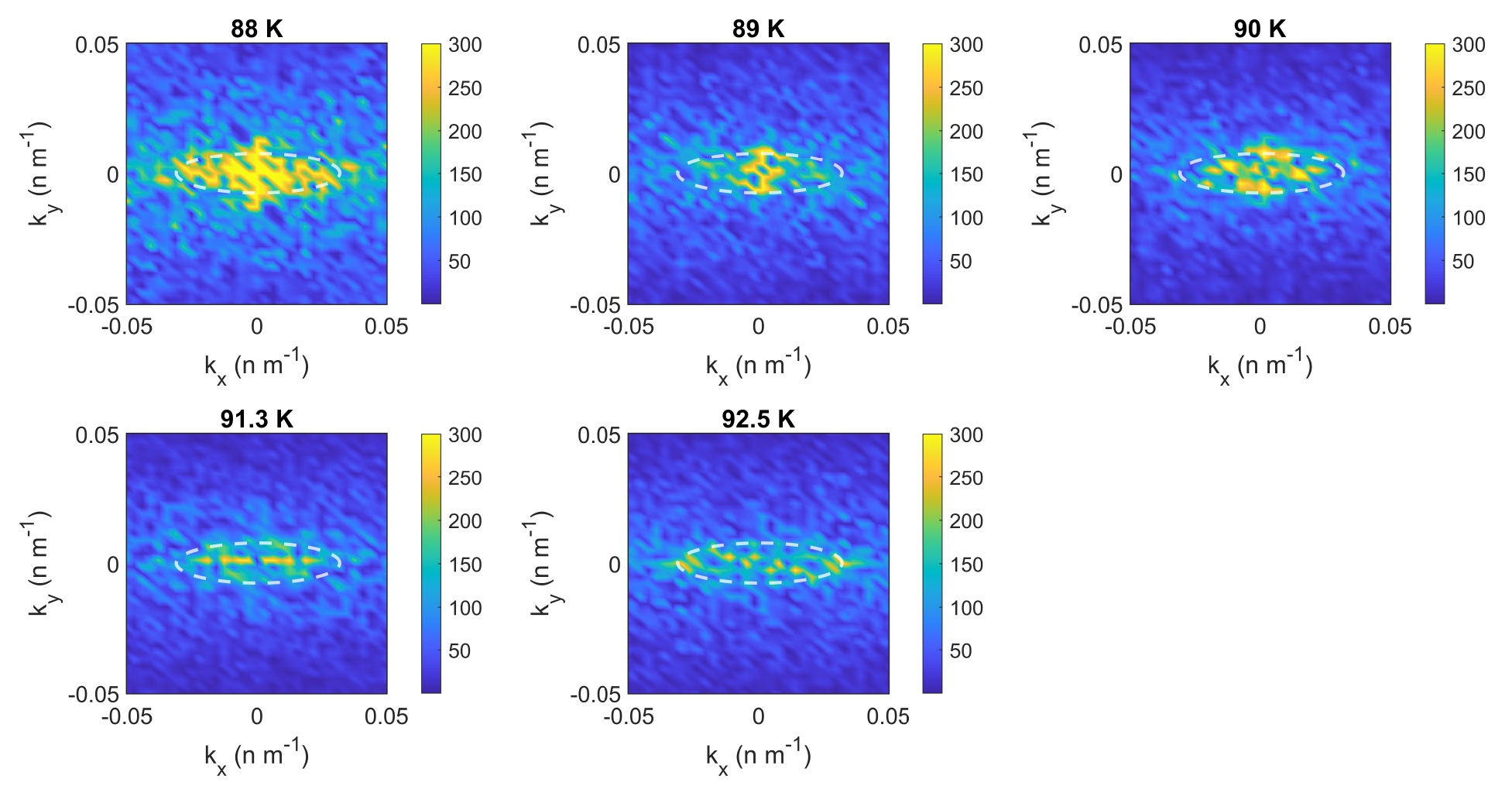}
\caption{The same LD PEEM data for FeSe plotted in Fig. \ref{FeSeFourier}, but with the magnitude clipped to highlight the ``spot" (dashed white line).}
\label{FeSeFourierclip}
\end{figure}

\begin{figure}[htb!]
\centering
\includegraphics[height=4in]{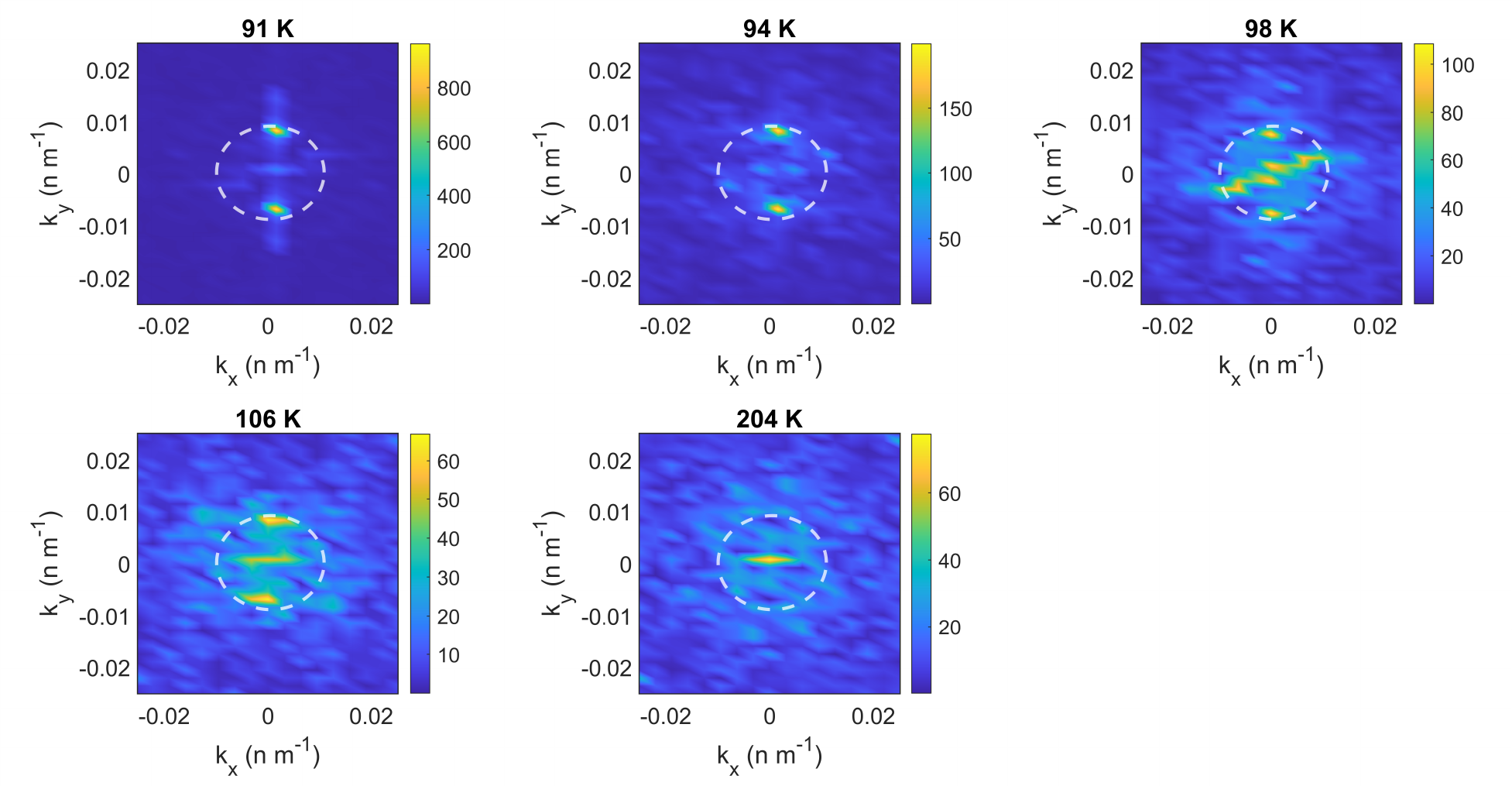}
\caption{The LD PEEM data for BaFe$_2$(As$_{0.87}$P$_{0.13}$)$_2$ in momentum space. The dashed white line, which is a guide to the eyes, marks the ``spot" of fixed size $\sim q_{\text{smc}}$. Such a spot is more clearly seen in Fig. \ref{BaFeAsPFourierclip}. This plot was generated from the real-space experimental data of Ref. \citen{Shimojima2021}, which is available at Zenodo  \cite{shimojima_takahiro_2021_4885407}.}
\label{BaFeAsPFourier}
\end{figure}

\begin{figure}[htb!]
\centering
\includegraphics[height=4in]{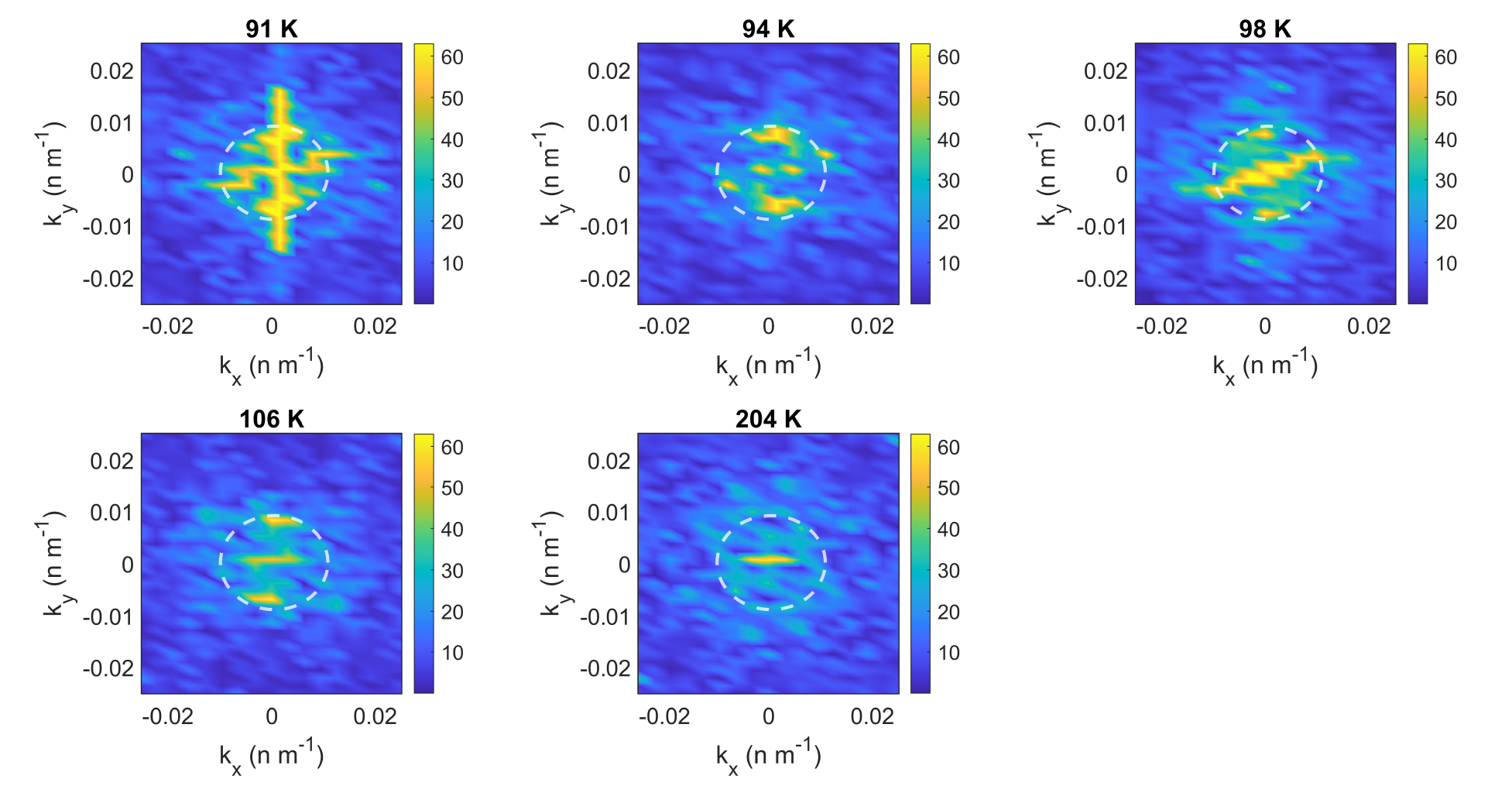}
\caption{The same LD PEEM data for BaFe$_2$(As$_{0.87}$P$_{0.13}$)$_2$ plotted in Fig. \ref{BaFeAsPFourier}, but with the magnitude clipped to highlight the ``spot" (dashed white line).}
\label{BaFeAsPFourierclip}
\end{figure}

We start by plotting the Fourier transform of the linear dichroism (LD) signal of the PEEM experiment \citep{Shimojima2021} in Figs. \ref{FeSeFourier}, \ref{FeSeFourierclip}, \ref{BaFeAsPFourier} and \ref{BaFeAsPFourierclip}. We note that the experimental data was extracted from Zenodo  \citep{shimojima_takahiro_2021_4885407}, where it is available, as stated in the manuscript where the data was originally published,  Ref. \citen{Shimojima2021}. 
Figs. \ref{FeSeFourier} and \ref{BaFeAsPFourier} show the Fourier-transformed data for FeSe and doped Ba-122 respectively, while Figs. \ref{FeSeFourierclip} and \ref{BaFeAsPFourierclip} show the same data with the intensity clipped for better visualization. In all cases, the momentum-space data displays a speckle pattern, from which we can infer the nematic fluctuations at non-zero momenta. Indeed, we can readily identify a ``spot" with an approximately fixed size, which we delineate by a dashed white ellipse in each panel for clarity (the spot is easier to visualize in Figs. \ref{FeSeFourierclip} and \ref{BaFeAsPFourierclip} since the intensity in Figs. \ref{FeSeFourier} and \ref{BaFeAsPFourier} is dominated by a pair of points). The radius of this spot corresponds to the experimentally observed $q_{\mathrm{smc}}$, and is present already at high temperatures. While the size of the spot changes only slightly as the temperature is lowered, the spectral weight is not only enhanced but also redistributed as $T$ decreases. In particular, at low enough temperatures, the spectral weight of the spot is concentrated at a sharply defined pair of spots along a specific direction, signaling the onset of a static smectic state.

This behavior is consistent with our theoretical description, from which we found that the defect-strains generate an effective smectic potential, $V_{\bm{q}_\parallel,z,z'}$. As shown in Fig. \ref{anisoptdef}, this potential displays sharp spots at the smectic wave-vector $q_{\text{smc}}$. This is consistent with the data in Figs. \ref{FeSeFourier} and \ref{BaFeAsPFourier} at the lowest temperatures, where smectic order sets in. Above the smectic transition temperature, the spectral weight is distributed inside an ellipse with radius of the order of $q_{\text{smc}}$, signaling the build-up of smectic fluctuations.




From the data in Fig. \ref{FeSeFourier}, we can roughly estimate from the distance between the sharp spots at low temperatures a smectic wave-vector of $q_\mathrm{smc} \sim 1 \times 10^{-2} \mathrm{nm}^{-1}$, corresponding to a wavelength of the order of hundreds of nanometers. Note that, according to Ref.  \citen{Shimojima2021}, the typical smectic domain size is of order of tens of microns.

%

We now perform a quantitative analysis to estimate the defect strength $\sigma$ required to obtain the experimentally determined value for $q_\mathrm{smc}$. Our goal is to determine whether this analysis gives a reasonable value for the defect strength -- alternatively, it can be compared to microscopic calculations, which are however beyond the scope of this work. We focus on FeSe, for which many of the material properties needed are known from other experiments. First, we construct the energy scale (per defect) associated with the defect distribution, $E_d=(1/N_{\text{step}})(h/\sigma)^2$, where $h$ is the average force exerted by each step defect. Using Eq. \ref{eq:aux1} for the smectic critical temperature, $r_0 (T_{\text{smc}}-T_{\text{nem}})/T_{\text{nem}}^{(0)}=b_\parallel q_{\text{smc}}^2= [(g\sigma)^4\beta^2/4b_\parallel][(\nu-1/2)^2+\nu^2]^2$, as well as the parameter $\beta$ defined in the main text, $\beta=[(1+\nu)/(\sqrt{2}E)]^{2}N_{\text{step}}(L_\xi/a_\parallel)$, we can express $E_d$ as:
\begin{align}
E_d=&\Bigg(\frac{g^4h^4[(\nu-1/2)^2+\nu^2]^2\big(\frac{1+\nu}{E\sqrt{2}}\big)^4 }{4 b_\parallel \frac{r_0 (T_{\text{smc}}-T_{\text{nem}})}{T_{\text{nem}}^{(0)}}}\Bigg)^{\frac{1}{2}} \\ \nonumber
= & \frac{g^2h^2[(\nu-1/2)^2+\nu^2]\big(\frac{1+\nu}{E\sqrt{2}}\big)^2 }{2 b_\parallel q_{\text{smc}} }. \label{eq:estimate}
\end{align}
Here, we approximated $L_\xi/a_\parallel \approx 1$ for simplicity.

In order to obtain an order-of-magnitude estimate for $E_d$, we need to estimate the various parameters that appear in Eq. (\ref{eq:estimate}): $g$, $b_{\parallel}$, $h$, and the elastic parameters $E$ (the Young modulus) and $\nu$ (the Poisson ratio). We will use the experimentally-determined value for the smectic wave-vector, $q_\mathrm{smc} \sim 1 \times 10^{-2} \mathrm{nm}^{-1}$. 

To obtain $g$ and $b_\parallel$, we need an estimate of $r_0$, which sets the energy scale of the inverse nematic susceptibility. Considering the nematic order's electronic origin, a reasonable estimate is $r_0\sim 1/N_F\sim E_F$, where $N_F$ is the density of states at the Fermi level and $E_F$ is the Fermi energy \citep{PhysRevResearch.2.013336}. The latter can be approximated by the Fermi energy for the hole band of FeSe, $E_F\sim 20$meV \citep{Shimojima2021,PhysRevB.96.121103,Chibani2021}.

As for the elastic parameters, firstly, we use the elastic constants reported in Refs. \citen{doi:10.1143/JPSJ.81.024604,CHANDRA20102072} to estimate the $B_{1g}$ elastic constant $C_{B_{1g}}=(C_{11}-C_{12})/\sqrt{2}\sim 20$GPa. Note that Ref. \citen{doi:10.1143/JPSJ.81.024604} considers the 2-Fe unit cell, (the puckered arrangement of the chalcogen/pnictogen atoms leads to two inequivalent Fe atoms in the Fe plane, as depicted in, say, Ref. \citen{PhysRevB.86.075123}), where the $x,y$ coordinate axes are aligned along the line joining two equivalent Fe atoms. In this situation, the structural distortion associated with nematicity occurs in the $B_{2g}$ channel, corresponding to the softening elastic constant $C_{66}$. Theoretical models often consider a 1-Fe unit cell (see, for instance, Ref. \citen{PhysRevB.86.075123}) for simplicity, with the $x,y$ coordinate axes aligned along the line joining two inequivalent Fe atoms, which is rotated relative to the one employed in the 2-Fe unit cell by $45^\circ$. In this case, the same nematic structural distortion happens in the $B_{1g}$ channel. Hence, for the present discussion based on the Ginzburg-Landau formalism, the shear modulus $C_{66}$ presented in Ref. \citen{doi:10.1143/JPSJ.81.024604} plays the role of our $C_{B_{1g}}$. Secondly, the Young's modulus and the Poisson's ratio are estimated as $E \sim 65$GPa, and $\nu \sim 0.2$, respectively \citep{CHANDRA20102072}. 

The nemato-elastic coupling $g$ can be obtained from the renormalization of the nematic critical temperature, for which we require not only the actual renormalized nematic critical temperature $T_{\text{nem}}\sim 90$K, but also the bare (i.e. not renormalized by the lattice) nematic transition $T_{\text{nem}}^{(0)}$. The latter has been obtained via Raman spectroscopy \citep{GALLAIS2016,Chibani2021}, which found $T_{\text{nem}}^{(0)}\sim 40$K. Using the relationship  $T_{\text{nem}}^{(0)}=T_{\text{nem}}-(g^2/C_{B_{1g}})(T_{\text{nem}}^{(0)}/r_0)$, we obtain $g\sim 1\times10^{-5}$J/m$^{3/2}$.
 
The nematic stiffness $b_\parallel$ can be obtained from the nematic correlation length $\xi$ via the relation
$\xi=\sqrt{b_\parallel T_{\text{nem}}^{(0)}/[r_0(T-T_{\text{nem}})]}$. The latter was obtained via inelastic X-ray scattering \citep{PhysRevLett.124.157001}, which reported $\xi(T=100K) \sim 70\mathrm{\AA}$, yielding $b_\parallel\sim 4\times10^{-38}$Jm$^2$.

Finally, while there is no data on $h$, the force exerted by each step, for iron-based superconductors, estimates for $h$ have indeed been obtained for simpler systems, yielding \citep{MULLER2004157} $h \sim 10^{-10}$N. For instance, Stewart et al. \citep{PhysRevB.49.13848} compare the displacement field measured by transmission electron microscopy to the corresponding analytical results on Si surfaces, obtaining $h=9.3\times 10^{-10}$N. Similarly, Shilkrot and Srolovitz \citep{PhysRevB.53.11120,PhysRevB.55.4737} obtain $h=2.9\times 10^{-10}$N on Ni and Au surfaces. We assume that the forces exerted by step defects in the systems considered in this study have similar magnitudes, and thus approximate $h\sim 10^{-10}$N. 

Substituting all the estimated parameters in Eq. (\ref{eq:estimate}), we find:
\begin{align}
E_d &\sim 100 \mu\text{eV}.
\end{align}
We emphasize that this value is just an order of magnitude estimate. 

\end{document}